\documentclass[
preprintnumbers,
preprint,
a4paper,
showpacs, 
amsmath, 
amssymb, 
floatfix,
nofootinbib,
]{revtex4}

\usepackage[dvips]{graphicx}

\begin{document}

\title{Black hole initial data in Gauss-Bonnet gravity:\\
Momentarily static case}

\author{Hirotaka Yoshino}

\affiliation{Cosmophysics Group, Institute of Particles and Nuclear Studies,
KEK, Tsukuba, Ibaraki, 305-0801, Japan}

\preprint{KEK-TH-1431}

\date{February 8, 2011}

%
%
\begin{abstract}
We study the method for generating the initial data
of black hole systems in Gauss-Bonnet (GB) gravity. The initial data
are assumed to be momentarily static and conformally flat.
Although the equation for the conformal factor is highly nonlinear,
it is successfully solved by numerical relaxation
for one-black-hole and two-black-hole systems.
The common apparent horizon is studied in the two-black-hole initial
data, and the result suggests that the Penrose inequalities
are satisfied in this system.
This is the first step for simulating
black hole collisions in higher-curvature theories.
\end{abstract}

\pacs{04.50.-h, 04.25.dg, 04.25.D-}
\maketitle

%
%
\section{Introduction}

Higher-dimensional gravity has been attracting a lot of attentions
motivated by the TeV-gravity scenarios \cite{ADD98-1,ADD98-2,RS99-1}.
If the three-dimensional space is a $3$-brane
in large or warped extra dimensions, the Planck energy
could be of $O({\rm TeV})$, and the trans-Planckian
collision will happen at accelerators such as the
Large Hadron Collider (LHC) \cite{BF99,DL01,GT02}. If this is the case,
the intermediate state of a collision with sufficiently
high energy and small impact parameter is expected to be a mini black hole,
and this motivated a lot of works (see \cite{Kanti:2009} for a review).
At this time, no evidence for black hole signals has been found,
and the restrictions for the Planck energy
and the minimum black hole mass $M_{\rm BH}^{\rm (min)}$
(such that the produced object can be regarded as a black hole 
if $M>M_{\rm BH}^{\rm (min)}$)
have been derived \cite{CMS:2010}.
Another motivation  for studying higher-dimensional gravity
is the AdS/CFT correspondence, which conjectures the
duality between gravity in anti-de Sitter (AdS) spacetime
and the conformal field theory (CFT) on the boundary of that spacetime.  
If this conjecture is correct, the phenomena in CFT
for which direct calculation is difficult due to
strong coupling effect 
can be predicted by calculating the dual gravitational system.

One of the important approaches for higher-dimensional gravity
is to explore the nonlinear dynamics of higher-dimensional spacetimes
by numerical relativity. Several formulations 
and codes of higher-dimensional numerical relativity
have been developed so far \cite{YS09,Zilhao:2010,Sorkin:2009}, 
and interesting simulations
have been performed: the time evolution of
Gregory-Laflamme instability \cite{Lehner:2010}, slow-velocity
collision of black holes \cite{Zilhao:2010,Witek1:2010,Witek2:2010}, 
dynamics of complex scalar field
minimally coupled to gravity in Kaluza-Klein spacetimes \cite{Sorkin:2009}, 
and bar-mode instability of rapidly rotating Myers-Perry
black holes \cite{SY09,SY10}. In these works, time evolutions of spacetimes are
simulated in the framework of 
general relativity (GR), which is
the simplest theory of higher-dimensional gravity. 

One of the interesting extensions of
higher-dimensional numerical relativity
is to include the higher-curvature terms.
In four dimensions, the Lagrangian density
that leads to the second-order equation
of the metric is just the Ricci scalar. However,
in higher-dimensions, the Lagrangian
density including higher-curvature
terms also leads to the second-order equation
if the combination of higher-curvature terms are chosen
appropriately. These theories are
Gauss-Bonnet (GB) gravity \cite{Lanczos} or, more generally,
Lovelock gravity \cite{Lovelock}. 
Among these higher-curvature theories,
we take attention to GB gravity in this paper.
The GB term gives a quadratic correction with respect to
the Riemann tensor to GR, and this GB correction 
arises in the low-energy 
limit of the heterotic string theory 
\cite{Zwiebach:1985,Duff:1986,Gross:1986}. Therefore, 
the GB term may have important effects 
on the mini black hole phenomena
at accelerators. Also, the AdS/CFT correspondence
in the context of GB gravity has been considered
(e.g., \cite{Ogushi:2004,Buchel:2009,Hu:2010}).

The black holes in GB gravity have interesting properties.
Although some part of GB gravity resembles GR, 
the other part does not. For example, the solution of a static
spherically symmetric black hole found in
Ref. \cite{BD85} has two branches: the non-GR branch
that is asymptotically AdS and the GR branch
that is asymptotically flat (see \cite{TM05-1,TM05-2} 
for a detailed study on causal structures
of these solution). The GB version of 
Birkoff's theorem \cite{Wiltshire:1988,Charmousis:2002}
(see also \cite{Zegers:2005} for Lovelock gravity) states that
the spherically-symmetric vacuum spacetime (at least locally)
corresponds to one of the two branches. The black hole
of GR branch is shown to be unstable for $D=5$ against
scalar mode and for $D=6$ against tensor mode if the
coupling constant $\alpha_{\rm GB}$ of the higher-curvature term
is sufficiently larger
than appropriate power of $GM$, where $G$ is the gravitational
constant and $M$ is the Arnowitt-Deser-Misner (ADM) mass of the black hole 
\cite{Dotti:2004,Dotti:2005,Gleiser:2005,Beroiz:2007}
(see also \cite{Takahashi:2010} for a study 
that proves the generic appearance
of instability of a spherically-symmetric black hole in 
Lovelock gravity). 

The GB version of numerical relativity (say, numerical
GB gravity), if it is developed,
would play important roles to explore nonlinear phenomena
in GB gravity. For example, the temporal evolution of the
instability of a small spherically symmetric black hole
could be followed by simulations to clarify the final end state.
Also, by simulating the time evolution from 
initial data of a nonstationary rotating black hole,
it would be possible to obtain an indication for the
stationary rotating black holes whose analytic solutions have not been
found to date (but see \cite{Brihaye:2010} for a numerical
construction of the black hole with two equal rotational parameters). 
Of course, the simulation of high-velocity collision of
black holes is an interesting issue to clarify the effect
of higher-curvature terms on mini black hole phenomena
at accelerators (see \cite{Rychkov:2004} for a partial result
for a head-on collision of the Aichelburg-Sexl particles in GB gravity).

There are few studies for numerical GB gravity. 
The $N+1$ formalism,
which is the extension of the ADM 
formalism to GB gravity, has been done by Torii
and Shinkai \cite{TS08}. Similarly to the ADM formalism, 
Einstein-Gauss-Bonnet equation is decomposed
into the Hamiltonian and momentum constraints
and the evolution equations. On the other hand,
the numerically stable formalism of GB gravity
which is analogous to, e.g., Baumgarte-Shapiro-Shibata-Nakamura (BSSN)
formalism \cite{Shibata:1995,Baumgarte:1998} has not been developed yet, and
also, no simulation of
numerical GB gravity has been done.

When one simulates a spacetime in GB gravity, similarly
to the case of GR, 
the first step is to prepare an initial spacelike hypersurface 
(i.e., initial data) by solving the constraint equations. 
In this paper, we focus attention to the
method for preparing initial data of
black hole systems in GB gravity. 
As a first step, the initial data are assumed
to be momentarily static (i.e., time symmetric).
In GR, the Brill-Lindquist initial data \cite{BL63}
are well known and widely used as the 
initial data of momentarily static multi black holes
(see also \cite{YN04,YSS05} for higher-dimensional studies). 
We discuss
the extension of the Brill-Lindquist initial data
to GB gravity, and successfully generate the
initial data for one black hole and two equal-mass black holes.

Using the two-black-hole initial data, we 
discuss whether the Penrose inequality \cite{Penrose} holds in this system.
The Penrose inequality states that the area of an apparent horizon (AH)
is not greater than that of the Schwarzschild-Tangherlini black hole
with the same ADM mass. 
In the case of the momentarily static initial data, 
the Penrose inequality was proved for $4\le D\le 7$ using the 
conformal flow method \cite{Bray:2007}. It is of interest
whether such a bound on the AH area holds
or not in GB gravity. In addition to the
original Penrose inequality, we also discuss
whether the AH area is bounded
from above by that of the spherically-symmetric
black hole in GB gravity. The results suggest that 
both of the inequalities are held in this system.

This paper is organized as follows. 
In the next section, we review GB gravity and $N+1$ formalism
in this theory focusing attention to the part that is related
to our study. In Sec.~III, general framework for constructing
the GB version of the Brill-Lindquist initial data
is explained. The equation for the conformal factor $\Psi$
is a Laplace equation with formal source term which is
highly nonlinear in $\Psi$, and we 
prove the regularity of the source term
which is necessary for the existence of the solution.
In Sec.~IV, the one-black-hole initial data are
constructed numerically, and we show the
agreement of the data with the time-symmetric slice 
of a spherically symmetric black hole spacetime. 
In Sec.~V, the two-black-hole initial data
are constructed. Using those data, we analyze the condition
for the AH formation, and the Penrose inequalities
in this system are discussed. Sec.~VI is devoted to summary
and discussion. 
Throughout this paper, we use the unit $c=1$, while
the gravitational constant $G$ is explicitly written.

%
%
\section{Constraint equations in Gauss-Bonnet gravity}

In this section, we review GB gravity and
the $N+1$ formalism of Ref.~\cite{TS08} focusing attention
to the part that is related to the setup of our study.

\subsection{Gauss-Bonnet action and equations}

In this paper, $D$ denotes the dimensionality of the spacetime $\mathcal{M}$
with the metric $g_{\mu\nu}$.
We also introduce $N = D-1$, which is the dimensionality
of a spacelike hypersurface $\Sigma$ in the spacetime $\mathcal{M}$. 
Throughout this paper, we assume the spacetime to be vacuum.
The Einstein-Gauss-Bonnet action \cite{Lanczos} in $D$-dimensional spacetime
is 
\begin{equation}
S=\frac{1}{16\pi G}\int_{\mathcal{M}}
\left(\mathcal{R}+\alpha_{\rm GB}
\mathcal{L}_{\rm GB}\right)\sqrt{-g}d^Dx
\label{action}
\end{equation}
with 
\begin{equation}
\mathcal{L}_{\rm GB} = \mathcal{R}^2
-4\mathcal{R}_{\mu\nu}\mathcal{R}^{\mu\nu}
+\mathcal{R}_{\mu\nu\rho\sigma}\mathcal{R}^{\mu\nu\rho\sigma},
\end{equation}
where $\mathcal{R}$, $\mathcal{R}_{\mu\nu}$, and $\mathcal{R}_{\mu\nu\rho\sigma}$
are the Ricci scalar, the Ricci tensor, and 
the Riemann tensor, respectively, and $\alpha_{\rm GB}$ is the coupling
constant with dimensionality of squared length.
The action~\eqref{action} gives the gravitational equation as
\begin{equation}
\mathcal{G}_{\mu\nu}+\alpha_{\rm GB}\mathcal{H}_{\mu\nu}=0,
\end{equation}
where
\begin{equation}
\mathcal{G}_{\mu\nu} = \mathcal{R}_{\mu\nu}
-\frac{1}{2}\mathcal{R}g_{\mu\nu},
\end{equation}
and
\begin{equation}
\mathcal{H}_{\mu\nu}=2\left(
\mathcal{R}\mathcal{R}_{\mu\nu}
-2\mathcal{R}_{\mu\alpha}\mathcal{R}^{\alpha}_{~\nu}
-2\mathcal{R}^{\alpha\beta}\mathcal{R}_{\mu\alpha\nu\beta}
+\mathcal{R}_{\mu}^{~\alpha\beta\gamma}\mathcal{R}_{\nu\alpha\beta\gamma}
\right)
-\frac{1}{2}g_{\mu\nu}\mathcal{L}_{\rm GB}.
\end{equation}
In the case $D=4$, 
the term $\mathcal{L}_{\rm GB}$ in the
action \eqref{action} gives a topological invariant 
as proved in the generalized Gauss-Bonnet theorem \cite{Chern}, and hence
$\mathcal{H}_{\mu\nu}=0$. The GB term becomes nontrivial only for
higher dimensions, $D\ge 5$.

%
%
\subsection{The $N+1$ formalism}

Here, we review the initial value equations
in GB gravity based on the $N+1$ formalism
developed in Ref.~\cite{TS08}. 
We consider the spacelike hypersurface $\Sigma$ with
the induced metric $\gamma_{\mu\nu}$ and the extrinsic curvature
$K_{\mu\nu}$, and let $n^{\mu}$ be the future-directed timelike 
unit normal to the hypersurface. The projections of the gravitational 
equation $(\mathcal{G}_{\mu\nu}+\alpha_{\rm GB}\mathcal{H}_{\mu\nu})n^\mu n^\nu$,
$(\mathcal{G}_{\mu\nu}+\alpha_{\rm GB}\mathcal{H}_{\mu\nu})n^\mu \gamma^\nu_{~\rho}$,
$(\mathcal{G}_{\mu\nu}+\alpha_{\rm GB}\mathcal{H}_{\mu\nu})
\gamma^\mu_{~\rho}\gamma^\nu_{~\sigma}$
give the Hamiltonian constraint, the momentum constraint, and the evolution
equations, respectively. 
The initial data should be prepared so that they
satisfy the two constraints.
The Hamiltonian constraint is written as
\begin{equation}
M+\alpha_{\rm GB}(M^2-4M_{ab}M^{ab}+M_{abcd}M^{abcd}) = 0,
\end{equation}
where
\begin{equation}
M_{ijkl}=R_{ijkl}+(K_{ik}K_{jl}-K_{il}K_{jk}),
\end{equation}
and $M_{ij}=\gamma^{ab}M_{iajb}$ and $M=\gamma^{ab}M_{ab}$. 
The explicit form of the other equations can
be found in Ref.~\cite{TS08}.

%
%
\subsection{Conformal approach}

Hereafter, we focus our attention
to the momentarily static case, or in other words,
the time-symmetric case: $K_{ij}=0$. In this case, 
the momentum constraint is trivially satisfied, and 
$M_{ijkl}=R_{ijkl}$. 
We further assume the initial space $\Sigma$
to be conformally flat:
\begin{equation}
\gamma_{ij} = \Psi^{4/(N-2)}\hat{\gamma}_{ij},
\end{equation}
where $\hat{\gamma}_{ij}$ is the flat-space metric.
Note that the authors of Ref.~\cite{TS08}
treated $\hat{\gamma}_{ij}$ as an arbitrary metric, and
here we chose the very special case. Also, 
in Ref.~\cite{TS08}, the conformal factor was set
to be $\Psi^{2m}$, where $m$ is an arbitrary number.
Here $m=2/(N-2)$ is chosen, because in this case,
we can treat the problem as a 
natural generalization of the GR studies.

In this setup, the equation for the conformal factor 
is written as
\begin{equation}
\hat{D}_a\hat{D}^a\Psi
=\alpha_{\rm GB}\hat{S},
\end{equation}
where $\hat{D}_a$ is the covariant derivative with respect to 
the flat-space metric. Here, $\hat{S}$ is written as
\begin{multline}
\hat{S} = \frac{N-3}{(N-1)(N-2)}\Psi^{-\frac{N+2}{N-2}}
\Bigg\{
4(N-2)\left[(\hat{D}_a\hat{D}^a\Psi)^2
-(\hat{D}_a\hat{D}_b\Psi)(\hat{D}^a\hat{D}^b\Psi)\right]
\\
-8\Psi^{-1}(\hat{D}\Psi)^2\hat{D}_a\hat{D}^a\Psi
+8N\Psi^{-1}\hat{D}^a\Psi\hat{D}^b\Psi\hat{D}_a\hat{D}_b\Psi
-\frac{4N(N-1)}{N-2}\Psi^{-2}(\hat{D}\Psi)^4
\Bigg\}.
\end{multline}

\section{Method of initial data construction}

In this section, we give the formulation
for generating the initial data with $N_{\rm BH}$ black holes
in GB gravity.

\subsection{Decomposition of conformal factor}

Since we would like to obtain the initial data
of GB gravity as continuous extension of those of the
GR case, we decompose $\Psi$ as
\begin{equation}
\Psi = \Psi_0 + \alpha_{\rm GB} g.
\label{decomposition}
\end{equation}
Here, $\Psi_0$ is the solution in the GR case 
(i.e., $\hat{D}_a\hat{D}^a\Psi_0=0$), and the term $\alpha_{\rm GB}g$ represents
the deviation from the GR case in the presence of $\alpha_{\rm GB}\neq 0$.
The equation for $g$ becomes
\begin{equation}
\hat{D}_a\hat{D}^a g = \hat{S}
\label{Eq-g-original}
\end{equation}
with
\begin{equation}
\hat{S} = \frac{N-3}{(N-1)(N-2)}\Psi^{-\frac{N+2}{N-2}}
\sum_{n=0}^4\alpha_{\rm GB}^n s^{(n)},
\label{source-decompose}
\end{equation}
where
\begin{multline}
s^{(0)} = -4(N-2)(\hat{D}_a\hat{D}_b\Psi_0)(\hat{D}^a\hat{D}^b\Psi_0)
+8N\Psi^{-1}(\hat{D}_a\Psi_0)(\hat{D}_b\Psi_0)(\hat{D}^a\hat{D}^b\Psi_0)
\\
-\frac{4N(N-1)}{N-2}\Psi^{-2}(\hat{D}\Psi_0)^4,
\label{source0}
\end{multline}
\begin{multline}
s^{(1)} =
-8(N-2)(\hat{D}_a\hat{D}_b\Psi_0)(\hat{D}^a\hat{D}^b g)
-8\Psi^{-1}(\hat{D}\Psi_0)^2(\hat{D}_a\hat{D}^a g)
\\
+8N\Psi^{-1}\left[
(\hat{D}_a\Psi_0)(\hat{D}_b\Psi_0)(\hat{D}^a\hat{D}^b g)
+2(\hat{D}^a\Psi_0)(\hat{D}^bg)(\hat{D}_a\hat{D}_b\Psi_0)
\right]
\\
-\frac{16N(N-1)}{N-2}\Psi^{-2}
(\hat{D}\Psi_0)^2(\hat{D}_a\Psi_0)(\hat{D}^ag),
\label{source1}
\end{multline}
\begin{multline}
s^{(2)} = -4(N-2)\left[(\hat{D}_a\hat{D}_bg)(\hat{D}^a\hat{D}^bg)
-(\hat{D}^a\hat{D}_ag)^2\right]
-16\Psi^{-1}(\hat{D}_a\Psi_0)(\hat{D}^ag)(\hat{D}^b\hat{D}_bg)
\\
+8N\Psi^{-1}
\left[
(\hat{D}^ag)(\hat{D}^bg)(\hat{D}_a\hat{D}_b\Psi_0)
+2(\hat{D}^a\Psi_0)(\hat{D}^bg)(\hat{D}_a\hat{D}_bg)
\right]
\\
-\frac{8N(N-1)}{N-2}\Psi^{-2}
\left[
2(\hat{D}_a\Psi_0\hat{D}^a g)^2
+(\hat{D}\Psi_0)^2(\hat{D}g)^2
\right],
\label{source2}
\end{multline}
\begin{multline}
s^{(3)}= -8\Psi^{-1}(\hat{D}g)^2(\hat{D}^a\hat{D}_ag)
+8N\Psi^{-1}(\hat{D}^ag)(\hat{D}^bg)(\hat{D}_a\hat{D}_bg)
\\
-\frac{16N(N-1)}{N-2}\Psi^{-2}(\hat{D}_a\Psi_0)(\hat{D}^a g)(\hat{D}g)^2,
\label{source3}
\end{multline}
and
\begin{equation}
s^{(4)}=-\frac{4N(N-1)}{N-2}\Psi^{-2}(\hat{D}g)^4.
\label{source4}
\end{equation}
We call the right hand side $\hat{S}$ of Eq.~\eqref{Eq-g-original} 
the ``source term'' hereafter.

\subsection{Puncture initial data}

Here, we specify the GR solution $\Psi_0$
of a system with $N_{\rm BH}$ black holes.
For this purpose,  
we introduce the Cartesian coordinates
$(x^a)$ to the flat space
and specify $N_{\rm BH}$ points at which $\Psi_0$ diverges (i.e., punctures). 
The location of the $n$-th puncture is denoted by $x^a = \bar{x}_{(n)}^a$,
and we define
$x_{(n)}^a = x^a - \bar{x}_{(n)}^a$ and 
$R_{(n)} = \left|x_{(n)}^a\right|$. Then, 
the solution to the Laplace equation $\hat{D}_a\hat{D}^a\Psi_0=0$ 
is adopted as
\begin{equation}
\Psi_0 = 1 + \sum_{n=1}^{N_{\rm BH}} \psi_{(n)}, 
\label{Psi0-Np-punctures}
\end{equation}
with
\begin{equation}
 \psi_{(n)} 
= \frac{4\pi GM_0^{(n)}}{(N-1)\Omega_{N-1}R_{(n)}^{N-2}},
\end{equation}
where $\Omega_{N-1}$ denotes the area of a $(N-1)$-dimensional 
unit sphere and $M_0^{(n)}$ is a mass parameter for $n$-th
black hole. The Arnowitt-Deser-Misner (ADM) mass $M_0$ is 
given as $M_0=\sum_{n=1}^{N_{\rm BH}}M_0^{(n)}$. 
This solution was studied in Ref.~\cite{BL63} in the $D=4$ case
and is often called the Brill-Lindquist
initial data (see \cite{YN04, YSS05} for the
studies in higher-dimensional cases). This space
possesses the $N_{\rm BH}$ Einstein-Rosen bridges (say, throats)
and each puncture corresponds to the asymptotically flat region 
beyond each throat.

\subsection{Finiteness of the source term}

\label{Sec:sourceterm}

The conformal factor $\Psi_0$ of the Brill-Lindquist initial
data in the GR case diverges at each puncture
as $R_{(n)}^{-(N-2)}$. Then, a naive estimate
gives a severely divergent behavior 
of the source term $\hat{S}\sim R_{(n)}^{-(N+4)}$
at each puncture. If this is the case, $g$ also has to
diverge at each puncture, and then, the behavior of $\hat{S}$
would be further modified, implying further stronger 
divergence of $g$. This procedure would continue eternally,
and hence the solution $g$ would not exist.

However, this naive estimate is not correct, since
the cancellation of divergent
terms occurs. As a result, $\hat{S}$ becomes zero at 
the punctures, and hence the source term $\hat{S}$ is well behaved.
Let us check this in the following.

Assuming a regular behavior of $g$ at each puncture,
the functions $s^{(2)}$, $s^{(3)}$, 
and $s^{(4)}$ obviously become zero
at each puncture after multiplying the
factor $\Psi^{-(N+2)/(N-2)}$ in Eq.~\eqref{source-decompose}. Therefore, 
we focus our attention only to  $s^{(0)}$
and $s^{(1)}$. We substitute
Eq.~\eqref{Psi0-Np-punctures} into Eqs.~\eqref{source0} and \eqref{source1}, 
rewrite it with $R_{(n)}\psi_{(n)}^{\prime\prime} + (N-1)\psi_{(n)}^{\prime}=0$, 
and collect only terms for which divergence at the punctures is suspected.
For example, $\Psi^{-2}(D\Psi_0)^4$ from the third term of Eq.~\eqref{source0} 
is calculated as
\begin{eqnarray}
\Psi^{-2}(D\Psi_0)^4 &=&
\Psi^{-2}\sum_{k,l,m,n} 
\psi_{(k)}^{\prime}\psi_{(l)}^{\prime}\psi_{(m)}^{\prime}\psi_{(n)}^{\prime}
(\vec{n}_{(k)}\cdot\vec{n}_{(l)})(\vec{n}_{(m)}\cdot\vec{n}_{(n)})
\nonumber\\
&=&
\Psi^{-2}\left[\sum_{k}\psi_{(k)}^{\prime 4}
+4\sum_{k,l (k\neq l)} \psi_{(k)}^{\prime}\psi_{(l)}^{\prime 3}
(\vec{n}_{(k)}\cdot\vec{n}_{(l)})+O(\psi^{\prime 2})\right],
\end{eqnarray}
where $n_{(n)}^a = x_{(n)}^a/R_{(n)}$ is the unit vector 
and $(\vec{n}_{(k)}\cdot\vec{n}_{(l)})$ is the inner product.
Here, $O(\psi^{\prime 2})$ are the terms that become zero
at the punctures after multiplying the factor 
$\Psi^{-(N+2)/(N-2)}$ in Eq.~\eqref{source-decompose}, and thus, 
we do not consider these terms because we are interested in
cancellation of divergent terms. In this manner, 
$s^{(0)}$ and $s^{(1)}$ are calculated as
\begin{multline}
s^{(0)} =
-\frac{4N(N-1)}{N-2}\sum_k
\left(\frac{\psi_{(k)}^\prime}{R_{(k)}\Psi}\right)^2
\left[R_{(k)}\psi_{(k)}^{\prime}
+(N-2)\Psi\right]^2 
\\
+4N
\sum_{k,l (k\neq l)} \frac{\psi_{(k)}^\prime\psi_{(l)}^\prime}{\Psi}
\left\{
\left[R_{(k)}\psi_{(k)}^\prime+R_{(l)}\psi_{(l)}^\prime+(N-2)\Psi\right]
\frac{1-N(\vec{n}_{(k)}\cdot\vec{n}_{(l)})^2}{R_{(k)}R_{(l)}}
\right.
\\
\left.
-\frac{4(N-1)}{N-2}
\left(\frac{\psi_{(l)}^\prime}{\Psi}\right)
\left[R_{(l)}\psi_{(l)}^\prime+(N-2)\Psi\right]
\frac{(\vec{n}_{(k)}\cdot\vec{n}_{(l)})}{R_{(l)}}
\right\}+O(R_{(k)}^{-2}),
\end{multline}
\begin{multline}
s^{(1)} = -8\sum_{k}
\left[R_{(k)}\psi_{(k)}^\prime + (N-2)\Psi\right]
\left(\frac{\psi_{(k)}^\prime}{R_{(k)}\Psi}\right)
\\
\times
\left[\hat{D}^a\hat{D}_ag
+Nn_{(k)}^an_{(k)}^b\hat{D}_b\hat{D}_ag
+\frac{2N(N-1)}{N-2}
\left(\frac{\psi_{(k)}^\prime}{\Psi}\right)
n_{(k)}^a\hat{D}_ag
\right]
+O(R_{(k)}^{-2}).
\end{multline}
Now, we use the relation 
\begin{equation}
R_{(n)}\psi_{(n)}^{\prime} + (N-2)\psi_{(n)} = 0.
\label{psi-relation2}
\end{equation}
and find 
$s^{(0)}=O(R_{(n)}^{-4})$ and $s^{(1)} = O(R_{(n)}^{-3})$.
Therefore, $\hat{S}$ behaves as $\sim R_{(n)}^{N-2}$ at each puncture,
and no divergence of the source term occurs.
For this reason, we can expect the existence of the solution 
of $g$ that is regular at each puncture.

In Secs.~IV and V,  
we explicitly construct the numerical solution of $g$
for one black hole and two black holes, respectively,
assuming the regularity of $g$ at each puncture.
There, it turns out that this numerical method works well.
It is worth pointing out the relation between our method
and the method for generating non-time-symmetric
initial data of puncture-type boosted
black holes in GR developed by Brandt and Br\"ugmann \cite{BB97}.
In their formalism, the space is assumed to be
conformally flat, and the analytic solution
of the extrinsic curvature 
found by Bowen and York \cite{BY80} is used.
The conformal factor is decomposed as $\Psi=\Psi_0+\psi$,
where $\Psi_0$ is the conformal factor for the
Brill-Lindquist solution. The equation for $\psi$
becomes the Laplace equation with a formal source term
depending on $\Psi$. Here, the source term is shown to become
zero at each puncture, and hence, $\psi$ can be solved numerically
assuming the regularity at each puncture. 
Therefore, our method is very analogous to the
Brandt-Br\"ugmann formalism.

\subsection{ADM mass}
\label{Sec:ADM}

Here, we discuss how to calculate the total gravitational energy. 
The ADM mass $M$ is given by
\begin{equation}
M = -\frac{(N-1)}{4\pi (N-2)G} \int_{\mathcal{S}} \hat{D}_a\Psi dS^a,
\end{equation}
where $\mathcal{S}$ is the surface at infinity.
Suppose $\Psi = \Psi_0$ is the conformal factor with the ADM mass $M_0$
in the GR case.
Then, using the Gauss law and Eq.~\eqref{Eq-g-original}, 
we find that the ADM mass in the GB case is 
\begin{equation}
M = M_0 -
\frac{(N-1)\alpha_{\rm GB}}{4\pi (N-2)G}\int_{\hat{\Sigma}}\hat{S}dx_1...dx_N,
\label{ADM_mass}
\end{equation}
where $\hat{\Sigma}$ is the whole flat space.
It is important to point out that the ADM mass $M$ in the
case $\alpha_{\rm GB}\neq 0$ is different from $M_0$,
although $M_0$ is the ADM mass in the GR case $\alpha_{\rm GB}=0$.
Therefore the parameter $M_0$ is (say) an artificial mass
for $\alpha_{\rm GB}\neq 0$,
and the true ADM mass $M$ is determined after $\Psi$ is solved.
The similar phenomena can be found also in the Brandt-Br\"ugmann formalism.

%
%
\section{One-black-hole initial data}

Let us begin our numerical analysis with the one-black-hole initial data.
Here we assume the initial data to be spherically 
symmetric and introduce
the radial coordinate $R$ 
in which the metric of the flat space becomes
\begin{equation}
ds^2 = dR^2 + R^2d\Omega_{N-1}^2.
\end{equation}
Here, $d\Omega_{N-1}^2$ is the line element of the
$(N-1)$-dimensional unit sphere. We  
impose $g=g(R)$  and set the function $\Psi_0$ in Eq.~\eqref{decomposition}
to be 
\begin{equation}
\Psi_0 = 1 + \left(\frac{R_S(M_0)}{R}\right)^{N-2},
\end{equation}
where we defined $R_S(M_0)$ as
\begin{equation}
R_S(M_0) := \left[\frac{4\pi GM_0}{(N-1)\Omega_{N-1}}\right]^{1/(N-2)}.
\label{def:R_S}
\end{equation}
Note that $R_S(M_0)$ is different from the
Schwarzschild radius $r_S(M_0)$, and they are related as
$r_S(M_0) = 4^{1/(N-2)}R_S(M_0)$. 

In the case of $\alpha_{\rm GB} = 0$, the conformal factor
$\Psi = \Psi_0$ gives the spacelike hypersurface which
is known as the Einstein-Rosen bridge (i.e., the time-symmetric Cauchy
slice in the Schwarzschild-Tangherlini spacetime).
The coordinate $R$ is called the isotropic coordinate, and
the minimal surface (or the apparent/event horizon)
is located at $R=R_S(M_0)$. Because of the GB version
of Birkoff's theorem \cite{Wiltshire:1988,Charmousis:2002}, also 
in the case of $\alpha_{\rm GB} > 0$, 
the initial data is expected to give the 
time-symmetric Cauchy slice of the spherically-symmetric black hole
spacetime \cite{BD85} whose metric is
\begin{equation}
ds^2 = -f(r)dt^2 +\frac{dr^2}{f(r)}+r^2d\Omega_{N-1}^2,
\label{spherically-symmetric1}
\end{equation}
with
\begin{equation}
f(r) = 1+\frac{r^2}{2\tilde{\alpha}_{\rm GB}}
\left(
1-\sqrt{1+\frac{4\tilde{\alpha}_{\rm GB}\tilde{M}}{r^{N}}}
\right),
\label{spherically-symmetric2}
\end{equation}
where $\tilde{\alpha}_{\rm GB} = (N-2)(N-3)\alpha_{\rm GB}$
and $\tilde{M} = \left[r_S(M)\right]^{N-2}$. If we can find
the coordinate transformation from the Schwarzschild-like
coordinate $r$ to the isotropic coordinate $R$, the conformal
factor $\Psi(R)$ is obtained. However, since $f(r)$ has
a complicated form in the cases $\alpha_{\rm GB}\neq 0$, 
it seems impossible to
find the analytic formula for the coordinate transformation
in contrast to the GR case. Therefore, this problem 
has to be solved numerically. Besides, we can obtain a
lot of lessons for the numerical method from this problem
as we will see later.

%
%
\subsection{Perturbative analysis}

In the numerical calculation, $R_S(M_0)$ is adopted
as the unit of the length, and also in the following, 
we use the unit $R_S(M_0)=1$ unless explicitly specified.
Let us consider the case where the coupling
constant $\alpha_{\rm GB}$ is small, and thus,
the correction from the GR case can be treated
as a perturbation. In this case, 
$\Psi$ in Eq.~\eqref{Eq-g-original} should be replaced by $\Psi_0$, and 
the equation is reduced to 
\begin{equation}
g_{,RR}+\frac{N-1}{R}g_{,R}
=-4N(N-3)R^{-2}\Psi_{0,R}^2\Psi_0^{-3-\frac{4}{N-2}}.
\end{equation}
Here, the cancellation of divergent terms has occurred
in the right-hand side
because of the relation $\Psi_{0,R}+(N-2)\Psi_0/R = (N-2)/R$, and
as a result, the source term is $O(R^{N-2})$ in the neighborhood
of the puncture.
In the case of $D=5$ (i.e., $N=4$), this equation
can be solved analytically:
\begin{equation}
g=\frac{4\left(1+3R^2+R^4\right)}{3\left(1+R^2\right)^3}.
\label{perturbative}
\end{equation}
Thus we explicitly confirm the existence of the
solution that is regular at the origin $R=0$. This result
leads to the expectation that a regular solution of $g$
exists in more general cases including higher-order terms in $\alpha_{\rm GB}$. 

The function $g$ behaves as $\simeq (4/3)R^{-2}$ 
at the distant region, $R\gg 1$. 
This indicates that the ADM mass is shifted as
\begin{equation}
M = M_0 + \frac{2\pi\alpha_{\rm GB}}{G}.
\end{equation}
Therefore, we confirm the statement of Sec.~\ref{Sec:ADM}
explicitly:
If we impose the regularity of $g$ at the origin,
the function $g$ also contributes to the mass, and hence,
the ADM mass is determined after $g$ is solved.

%
%
\subsection{Numerical approach}

Now, we study the numerical generation of $g(R)$
for the general cases of $\alpha_{\rm GB}\ge 0$. 
We write down the Laplace operator $\hat{D}_a\hat{D}^a$
and the source term $\hat{S}$
in terms of the radial coordinate
$R$. Here, we adopt the formulas of $s^{(0)}$ and $s^{(1)}$
for which the divergent terms are canceled out by
the relation 
$\Psi_{0,R}+(N-2)\Psi/R = (N-2)(1+\alpha_{\rm GB}g)/R$.
Then, the right hand side
becomes $O(R^{N-2})$ similarly to the perturbative case
and the source term $\hat{S}$
is well behaved at the origin $R=0$.
This confirms the general proof for the finiteness
of $\hat{S}$ of Sec.~\ref{Sec:sourceterm}.

The numerical calculation is done in a finite region, and hence,
there is an outer boundary of the computation domain, $R=R_{\rm max}$.
Here, we have to specify the outer boundary condition. 
As discussed in the previous subsection, the behavior
of $g$ at $R\gg 1$ is expected to be $g\simeq C/R^{N-2}$. Since 
the value of $C$ is not known before solving $g$, we 
eliminate $C$ using the combination of $g_{,R}$ and $g$ as
\begin{equation}
g_{,R}+(N-2)g/R=0,
\label{BC:Robin}
\end{equation}
and adopt this Robin boundary condition
at $R=R_{\rm max}$.

We used the finite differencing method with the fourth-order
accuracy with respect to the grid size. Here,
the method of nested hierarchical grids
was adopted in the numerical calculation. Specifically,
we located the outer boundary at $R_{\rm max}=1024$,
and put 12 and 16 layers for $0\le\alpha_{\rm GB}< 10^2$
and $10^2\le \alpha_{\rm GB}$, respectively 
(remember that the unit of the length is $R_S(M_0)$).
The boundary of the $n$-th layer is located at $R=1024/2^{n-1}$,
and the grid size is $12.8/2^{n-1}$. Then, the solutions were
obtained by using the successive-over-relaxation (SOR) method:
For each calculation, a test surface is prepared initially and 
it is made slowly converge to the real solution until
the difference from the finite difference equations normalized
by the absolute value of $g$ becomes less than $10^{-12}$.

%
\begin{figure}[tb]
\centering
{
\includegraphics[width=0.45\textwidth]{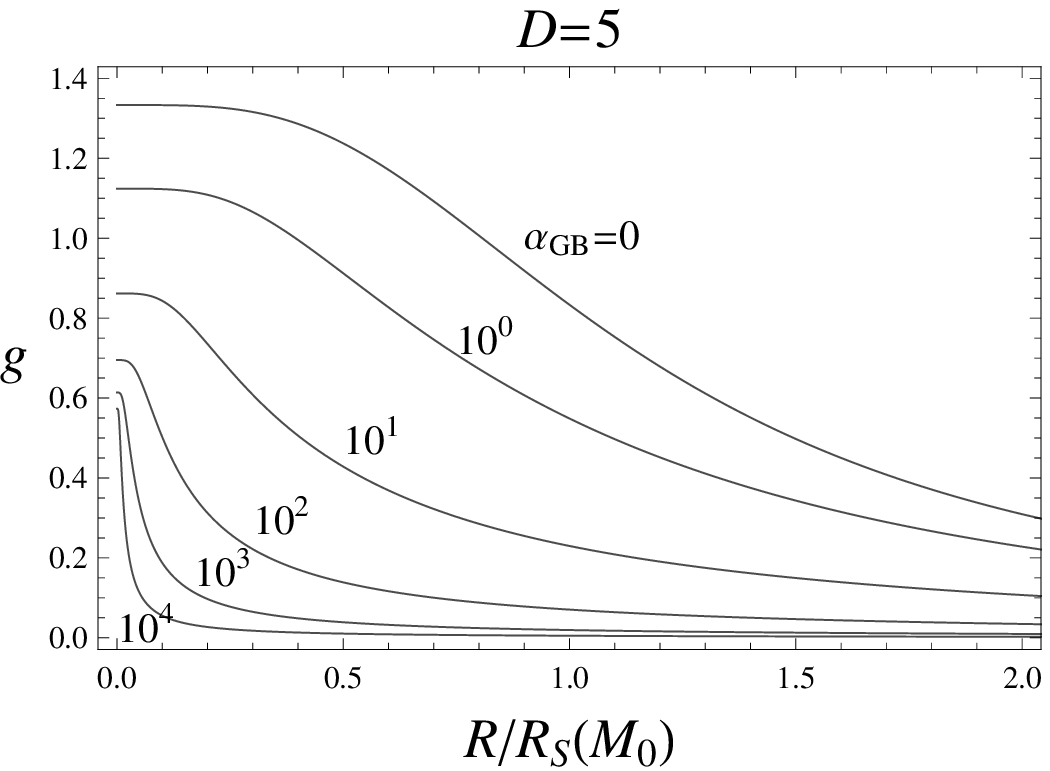}
\includegraphics[width=0.45\textwidth]{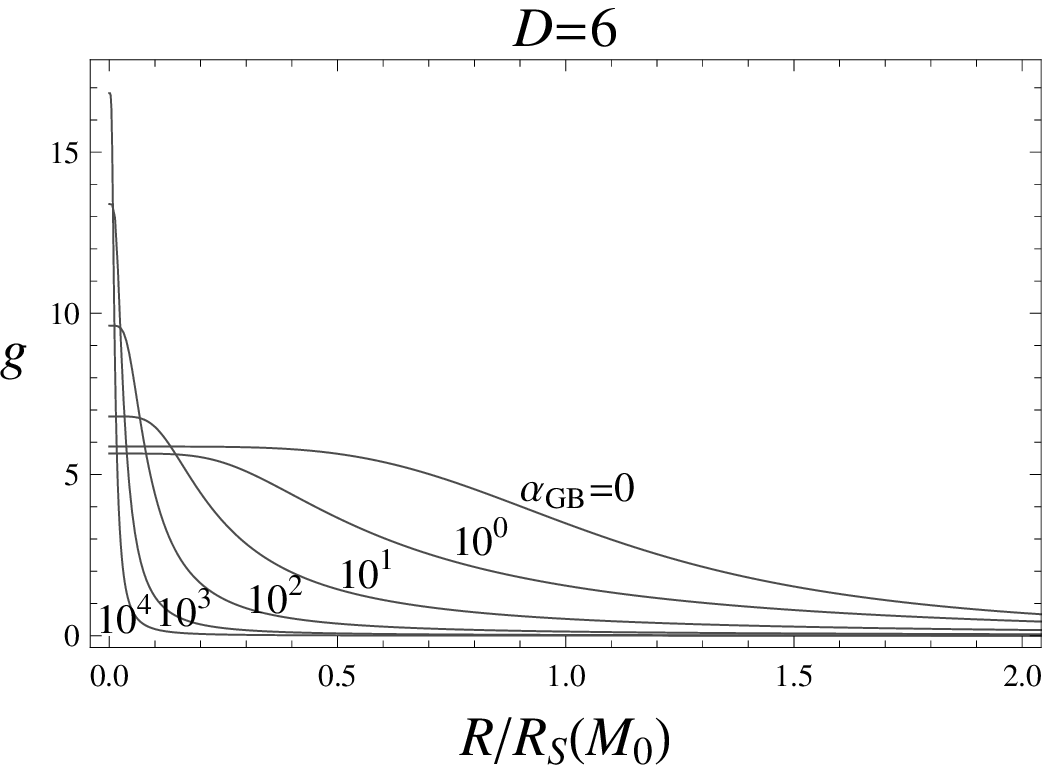}
\includegraphics[width=0.45\textwidth]{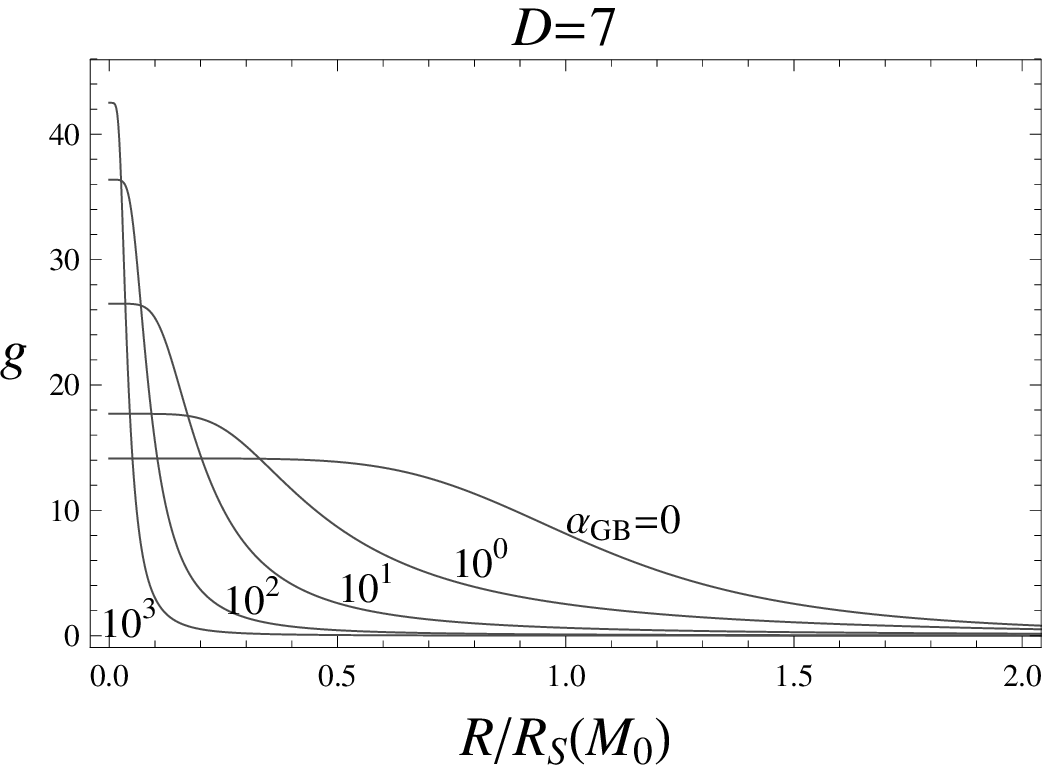}
\includegraphics[width=0.45\textwidth]{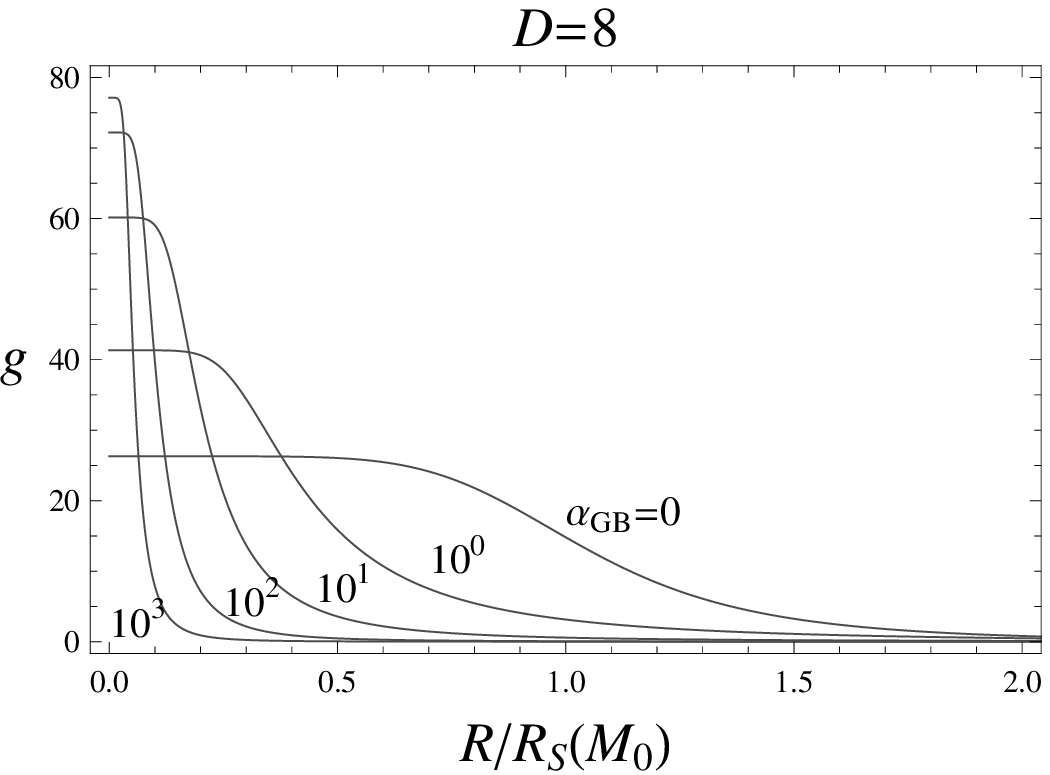}
}
\caption{The behavior of $g$ (in the unit of $R_S^2(M_0)$)
as functions of $R/R_S(M_0)$ for $D=5$, $6$, $7$, and $8$
for the cases $\alpha = 0$ and $10^k$ with $0\le k\le 3$.
For $D=5$ and $6$, the case $\alpha=10^4$ is also shown. 
}
\label{g-behave}
\end{figure}
%

Figure~\ref{g-behave} shows the behavior of $g$
as a function of $R$ for $\alpha_{\rm GB}=0$ and $10^k$
for $D=5$--$8$, where $k=1,...,4$
for $D=5$ and $6$ and $k=1,...,3$ for $D=7$ and $8$. 
The curve for $\alpha_{\rm GB}=0$ in the case $D=5$
agrees with Eq.~\eqref{perturbative}. 
As $\alpha_{\rm GB}$ is increased, the curve becomes steeper
around the origin. This is the reason why we increased the
number of the layers for $100\le \alpha_{\rm GB}$: More grid number
is required for larger $\alpha_{\rm GB}$.

%
\begin{figure}[tb]
\centering
{
\includegraphics[width=0.45\textwidth]{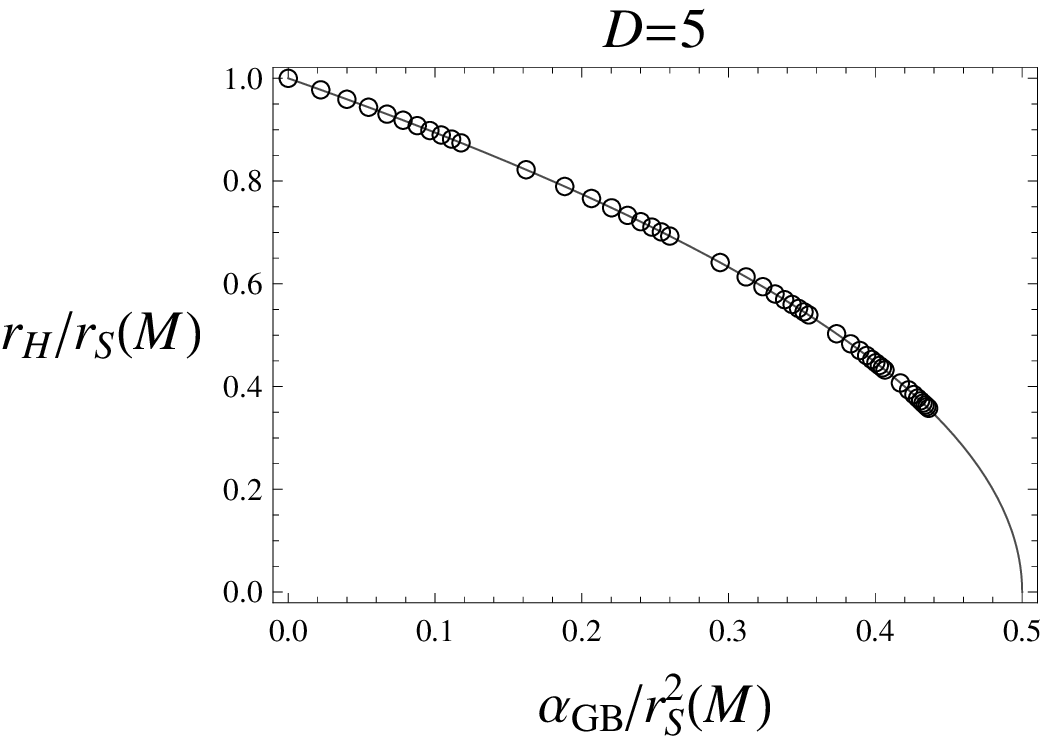}
\includegraphics[width=0.45\textwidth]{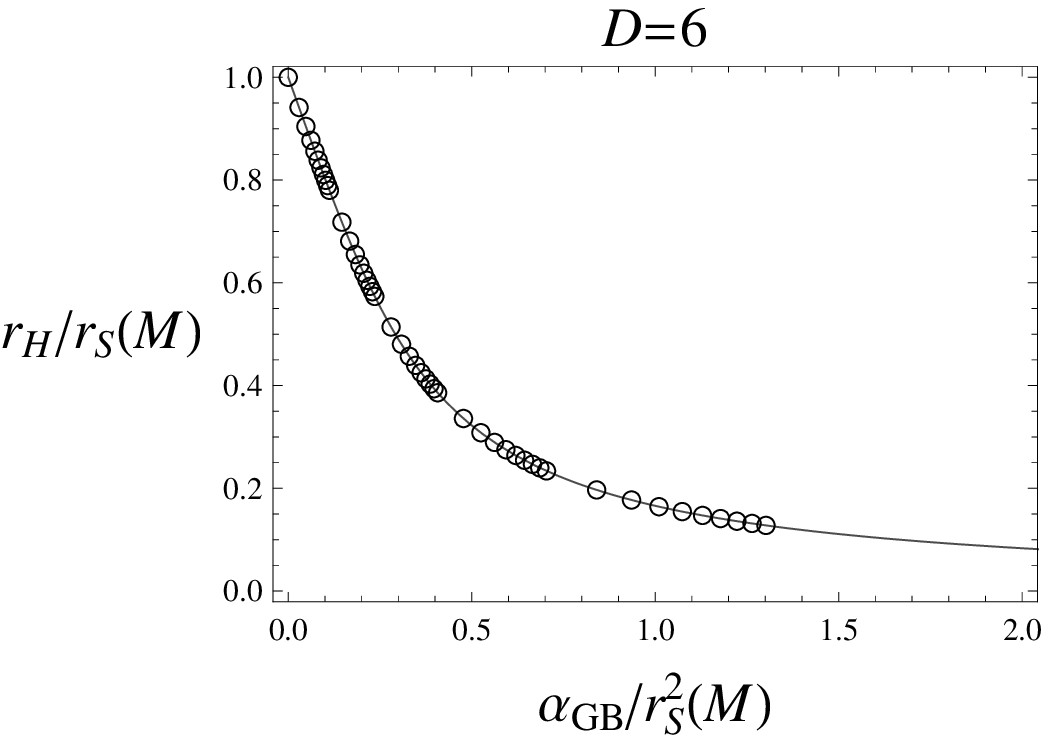}
\includegraphics[width=0.45\textwidth]{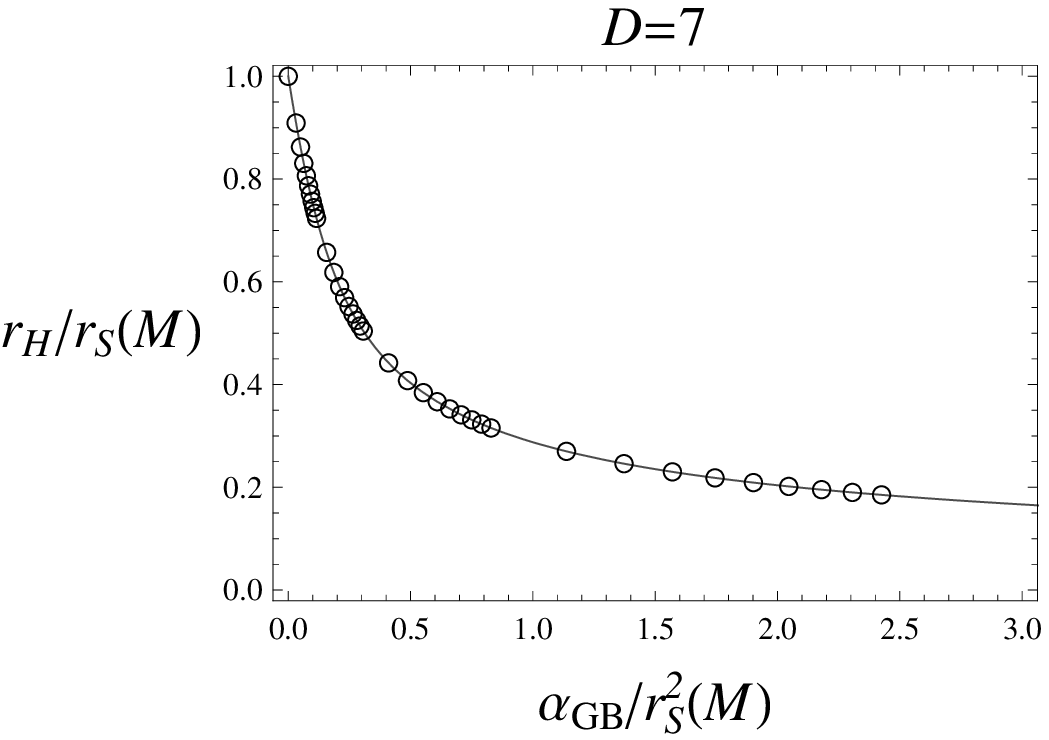}
\includegraphics[width=0.45\textwidth]{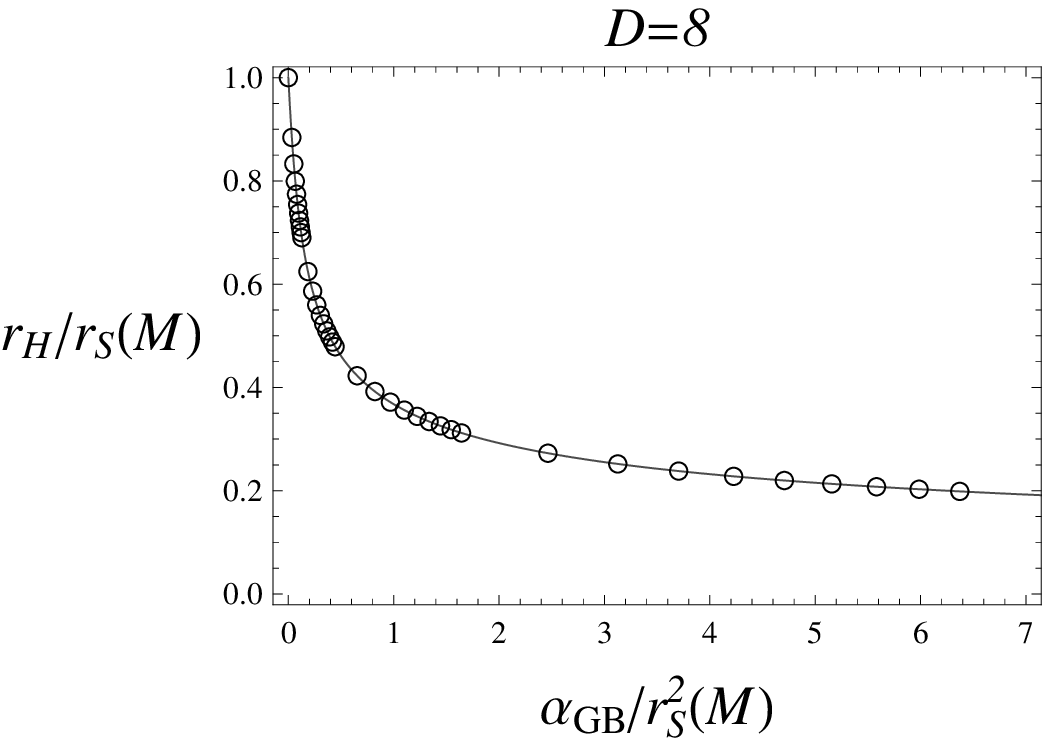}
}
\caption{The radius $r_H/r_S(M)$ of the minimal surface as a
function of $\alpha_{\rm GB}/r_S^2(M)$ for $D=5$, $6$, $7$, and $8$. 
Each straight curve shows the analytic relation, 
while the circles ($\circ$) show the numerical result.
The numerical data are taken for $\alpha_{\rm GB}/R_S^2(M_0) =0$, 
$n\times 10^{k-1}$ 
(where $n$ and $k$ are integers satisfying $1\le n\le 9$ 
and $0\le k\le k_{\rm max}$), and $10^{k_{\rm max}}$,
where $k_{\rm max}=4$ for $D=5$ and $6$ whereas $k_{\rm max} = 3$
for $D=7$ and $8$.
The data agree well with the analytic relation.}
\label{alpha_radius}
\end{figure}
%

In order to check that the geometry of the generated initial data agrees
with that of the time-symmetric Cauchy surface in the spherically-symmetric
spacetime, Eqs.~\eqref{spherically-symmetric1} and
\eqref{spherically-symmetric2}, we calculate the minimal surface
(or the AH) and compare the relations between
the two nondimensional quantities,
$\alpha_{\rm GB}/r_S^2(M)$ and $r_H/r_S(M)$, where
$r_H$ is the horizon radius.
The horizon of the analytic solution is given by $f(r_H)=0$, which
leads to the relation
\begin{equation}
\left(\frac{r_H}{r_S}\right)^{N-4}
\left[\left(\frac{r_H}{r_S}\right)^{2}
+\frac{\tilde{\alpha}_{\rm GB}}{r_S^2}\right]=1.
\label{Eq:horizon_radius}
\end{equation}
On the other hand, we find the minimal surface of the numerical data
by searching the location $R=R_H$ at which 
\begin{equation}
\frac{2}{N-2}R\Psi_{,R}+\Psi=0
\end{equation}
is satisfied, and calculate the horizon radius as
\begin{equation}
r_H = R_H\Psi^{2/(N-2)}(R_H).
\end{equation}
Then, the ADM mass $M$ is calculated by Eq.~\eqref{ADM_mass}, and
the value of $r_H/r_S(M)$ is evaluated.

Figure~\ref{alpha_radius} shows the relation between 
$\alpha_{\rm GB}/r_S^2(M)$ and $r_H/r_S(M)$ for $D=5$--$8$.
The solid curve shows the analytic formula, and
the circles are our numerical data points.
They agree very well, and thus, we can confirm
that the generated initial data
is the time-symmetric slice in the spherically-symmetric spacetime. 
The cases for $\alpha_{\rm GB}/R_S^2(M_0)=0$, $n\times 10^{k-1}$ 
(where $n$ and $k$ are integers satisfying $1\le n\le 9$ 
and $0\le k\le k_{\rm max}$), and $10^{k_{\rm max}}$ 
are shown for all $D$. Here,
$k_{\rm max}$ is chosen as $k_{\rm max}=4$ for $D=5$ and $6$,
and $k_{\rm max}=3$ for $D=7$ and $8$.
As seen from this figure, the value of $r_H/r_S(M)$ is decreased
as $\alpha_{\rm GB}/R_S^2(M_0)$ is increased, but 
even at $\alpha_{\rm GB}/R_S^2(M_0)=10^4$, the value of $r_H/r_S(M)$
is $\simeq 0.36$ and $0.13$ for $D=5$ and $6$, respectively.
Therefore, generating the initial data with a horizon radius $r_H$
that is much smaller than the Schwarzschild radius $r_S$
requires a very large number of $\alpha_{\rm GB}/R_S^2(M_0)$,
and thus, it is a hard task at least in this approach.

%
\begin{figure}[tb]
\centering
{
\includegraphics[width=0.45\textwidth]{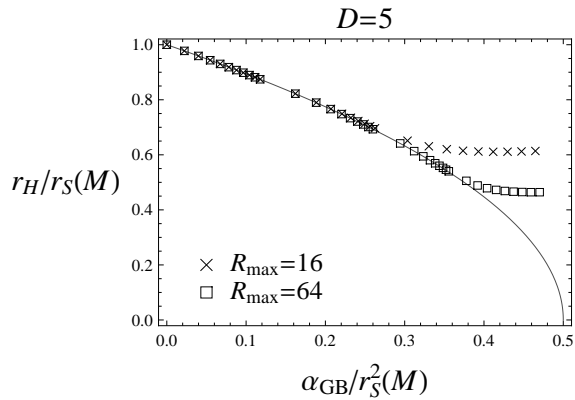}
}
\caption{The numerical data taken for $R_{\rm max}=16$ (crosses, $\times$)
and $64$ (squares, $\square$) in the case of $D=5$. 
The deviation from the analytic relation becomes
significant for $\alpha_{\rm GB}/R^2_S(M_0)\gtrsim 10$ and $100$ 
for $R_{\rm max}=16$ and $64$, respectively, 
indicating that the larger number of $R_{\rm max}$
is required for accurate numerical calculations as 
$\alpha_{\rm GB}$ is increased.
}
\label{boundaryerror}
\end{figure}
%

Here, we point out the importance of the location of the
outer boundary $R_{\rm max}$. Figure~\ref{boundaryerror}
shows the numerical results with the choices of
$R_{\rm max}=16$ and $64$. The agreement with the analytic
relation is fairly good for small $\alpha_{\rm GB}$, 
but the deviation becomes significant for
$\alpha_{\rm GB}/R_S^2(M_0)\gtrsim 10$ and $100$ for 
$R_{\rm max}=16$ and $64$, respectively. 
This is because the integrand of Eq.~\eqref{ADM_mass} 
cannot be ignored for $R>R_{\rm max}$ and therefore
the ADM mass $M$ is evaluated to be a smaller value
than the true value. We have to remember that
the larger value of $R_{\rm max}$ 
is required for a larger number of $\alpha_{\rm GB}$
for accurate calculations.

Now we summarize the lessons obtained in the analysis
in this section: (i) The source term for the equation $g$
becomes zero at $R=0$ because of the cancellation of divergent
terms, and thus, the regular solution can be constructed;
(ii) The ADM mass $M$ has to be evaluated after the function $g$ 
is generated, because it contributes to the mass;
(iii) The function $g$ becomes steeper around $R=0$ as
$\alpha_{\rm GB}$ is increased, and therefore, the better resolution
is required; and (iv) We have to take care of the location of the
outer boundary $R_{\rm max}$ for large $\alpha_{\rm GB}$ values 
since the integrand of Eq.~\eqref{ADM_mass} becomes large at the
distant region.

%
%
\section{Two-black-hole initial data}

Now we turn our attention to the study on the two-black-hole initial
data. After generating the conformal factor, we
study the common AH and assess whether
the Penrose-like inequalities are held in this system.

%
%
\subsection{Numerical calculation}

We assume the two black holes to have the same mass
and the initial space to be axisymmetric
with $O(N-1)$ symmetry (i.e., there is the
$z$-axis and the directions orthogonal to axis have the
same structure). To be specific, we introduce the
$(z, \rho)$ coordinates where the metric of the
flat space is
\begin{equation}
ds^2=dz^2 + d\rho^2+\rho^2 d\Omega_{N-2}^2,
\end{equation}
and assume $\Psi=\Psi(z,\rho)$.
The GR solution $\Psi_0$ 
is adopted as the Brill-Lindquist solution with two equal-mass
black holes that is described as
\begin{equation}
\Psi_0 = 1+\frac{1}{2}[R_S(M_0)]^{N-2}
\left(\frac{1}{R_+^{N-2}}
+\frac{1}{R_-^{N-2}}\right).
\end{equation}
Here, the punctures are located at $z=\pm z_0$ on the $z$ axis
and 
\begin{equation}
R_\pm := \sqrt{(z\mp z_0)^2+\rho^2},
\end{equation} 
and $R_S(M_0)$ is
defined in Eq.~\eqref{def:R_S}. In this setup, the space 
possesses the two throats and three asymptotically
flat regions (say, one upper sheet and two lower sheets).
The input parameters in the numerical calculations are 
$z_0$ and $\alpha_{\rm GB}$ in the unit $R_S(M_0)=1$.

In the numerical calculation of $g(z,\rho)$, we write down the Laplace
operator $\hat{D}_a\hat{D}^a$ and the source term $\hat{S}$
in terms of $(z, \rho)$, where the divergent terms
of Eqs.~\eqref{source0} and \eqref{source1} 
(i.e., $s^{(0)}$ and $s^{(1)}$) are canceled out.
Since the two black holes have the same mass, there is
a mirror symmetry with respect to $z=0$. For this reason,
we choose the computation domain as
$0\le z\le z_0+\Delta z_{\rm max}$ and $0\le \rho\le \Delta\rho_{\rm max}$
where $\Delta z_{\rm max}$ and $\Delta\rho_{\rm max}$
are chosen as $\Delta z_{\rm max} = \Delta \rho_{\rm max} =1024$
in the unit $R_S(M_0)=1$. 
At the outer boundary, we impose the same boundary condition
\eqref{BC:Robin} as the spherically symmetric case but
rewritten in the $(z, \rho)$ coordinates.
Similarly to the case of one-black-hole initial data,
we used the fourth-order finite differencing
and the method of nested hierarchical grids.
We put 13 layers to the computational domain,
where the $n$-th layer has the boundary
at $z=z_0 \pm \Delta z_{\rm max}/2^{n-1}$ and 
$\rho = \Delta \rho_{\rm max}/2^{n-1}$. 
If the layer crosses $z=0$, the region
$z<0$ is discarded. 
The grid number $N_{\rm grid}$ of $\rho$ coordinate  
of each layer is varied as $10$, $20$, and $40$ depending on the situation,
and the grids of $z$ coordinate have the same size.
Then, the solution is obtained by the SOR method.
The relaxation is continued until the deviation from the
finite difference equations normalized by the absolute value of $g$
becomes less than $10^{-12}$.

%
\begin{figure}[tb]
\centering
{
\includegraphics[width=0.45\textwidth]{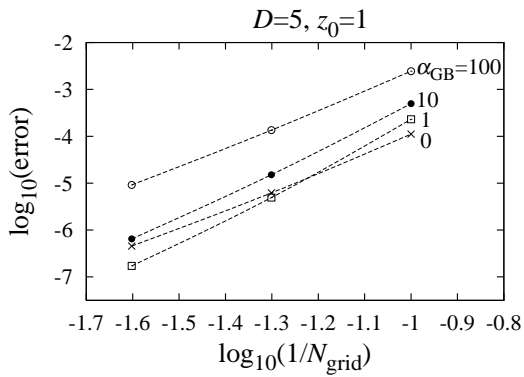}
}
\caption{Relation between numerical error in $g(z,\rho)$
and inverse of the grid number $(1/N_{\rm grid})$ (proportional
to the grid size) for $\alpha_{\rm GB}=0$ ($\times$), 
$1$ ($\square$), $10$ ($\bullet$), and $100$
($\circ$)
for the case $D=5$ and $z_0=1$.
}
\label{D5-error}
\end{figure}
%

The validity of the numerical computation is checked in three manners.
First, it is checked that the numerical solution $g(z,\rho)$ 
in the case of $z_0=0$ agrees with 
that of one-black-hole initial data.
Next, as $z_0$ is increased, the solution is expected
to asymptote to that of one isolated black hole
with half mass in the neighborhood of each puncture, and 
the numerical solution $g(z,\rho)$ is actually confirmed
to show this behavior. 
Finally, we checked whether the numerical solution shows
the appropriate convergence with respect to the grid size.
Figure~\ref{D5-error} shows the typical numerical error in $g(z,\rho)$
for $\alpha_{\rm GB}=0$, $1$, $10$, and $100$ 
for the case $D=5$ and $z_0=1$ as a function of $\log_{10}(1/N_{\rm grid})$.
Here, the error is evaluated for $N_{\rm grid}=10$, $20$, and $40$
by calculating the difference from the solution of the case $N_{\rm grid}=80$. 
The slope for the curves of $\alpha_{\rm GB}=0$ and $100$ is approximately
four, reflecting the fact that the adopted method is the
fourth-order accuracy scheme. On the other hand, the slope for the curves
of $\alpha=1$ and $10$ is larger than four; the slope for $\alpha=1$
is approximately five. This is a somewhat strange result, because
the five-order accuracy was obtained by using the scheme
with the fourth-order accuracy.
This is probably because the numerical error in $g$ changes the value of
the right-hand side $\hat{S}$ of Eq.~\eqref{Eq-g-original}, and the change in
$\hat{S}$ further modify the value of $g$. Because of this
effect, the cancellation of the numerical error would have
happened accidentally.
Anyway, the numerical data shows at least the fourth-order convergence,
and this supports the accuracy of our numerical calculation.

%
\begin{figure}[tb]
\centering
{
\includegraphics[width=0.3\textwidth]{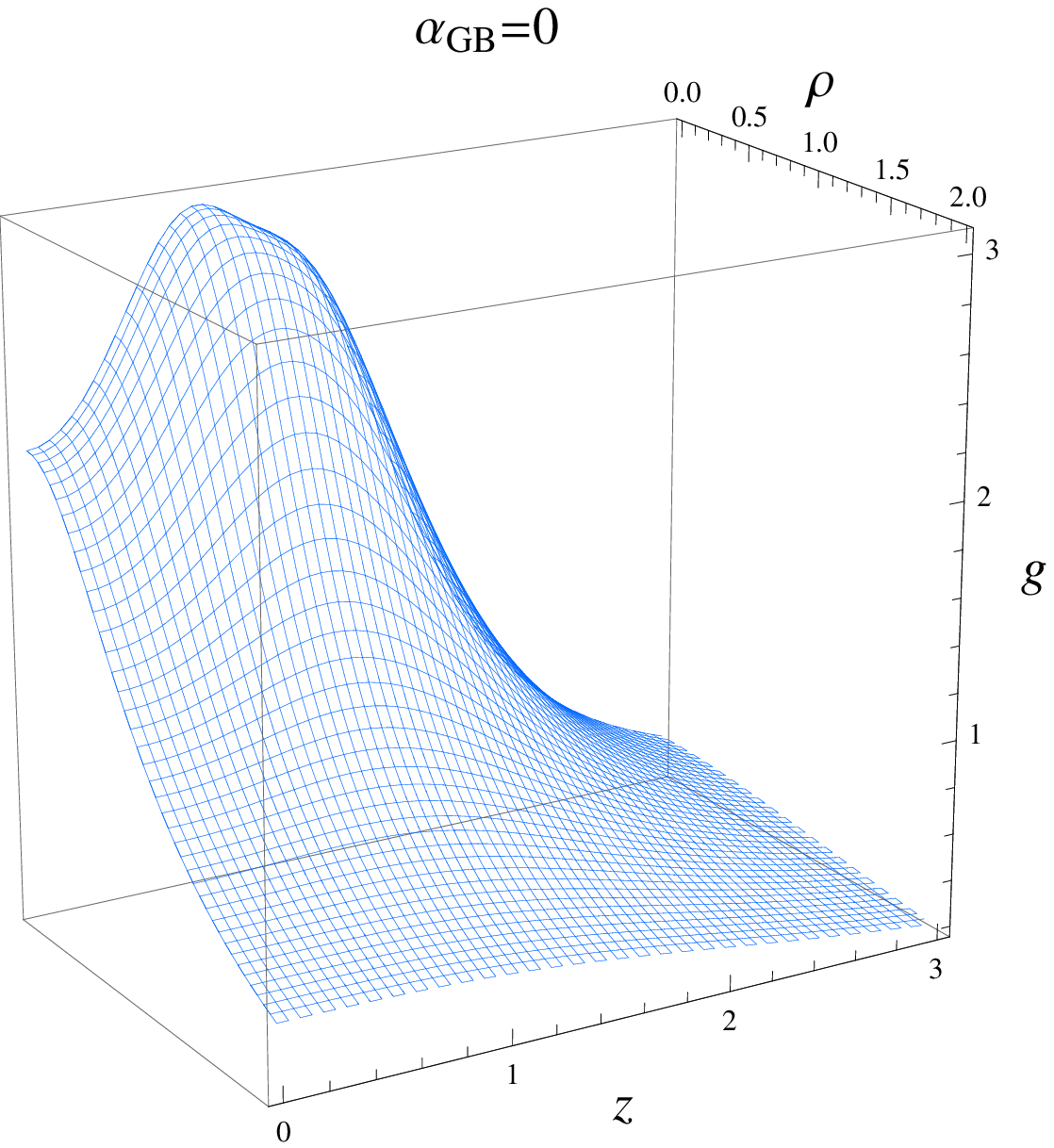}
\includegraphics[width=0.3\textwidth]{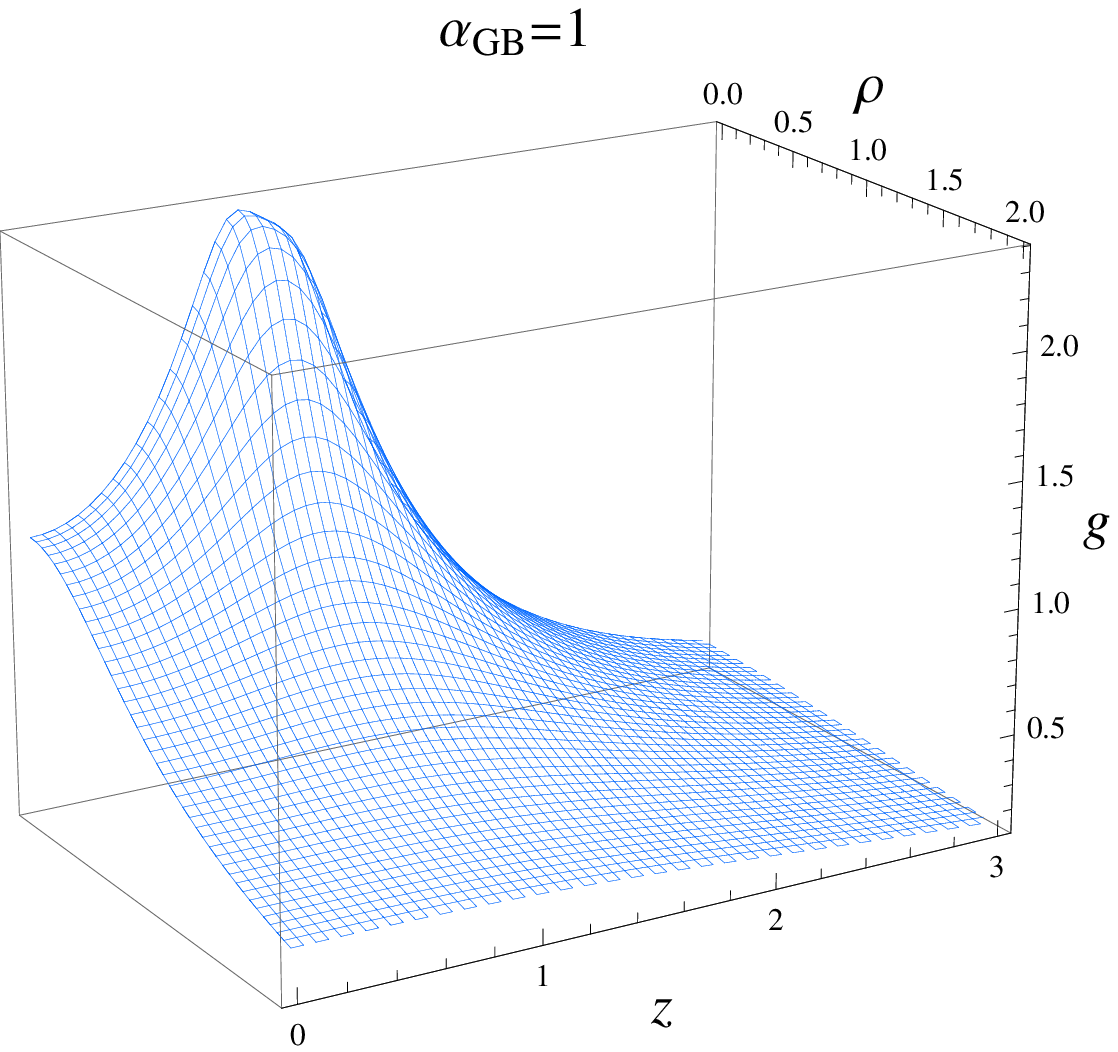}
\includegraphics[width=0.3\textwidth]{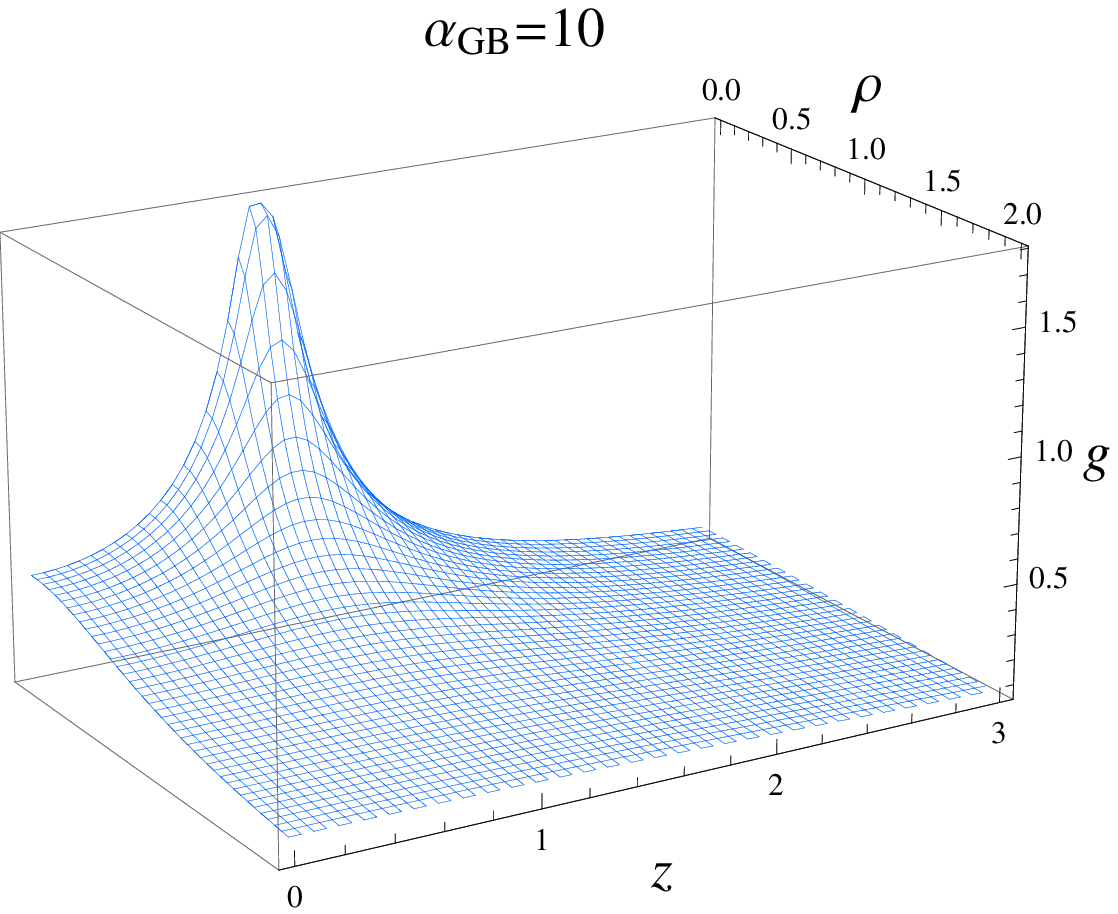}
}
\caption{3D plot of the function $g(z,\rho)$
in the domain $0\le z\le 3$ and $0\le \rho\le 2$
in the case $z_0=1$ and $D=5$. 
The cases for $\alpha_{\rm GB}=0$ (left), $1$ (center), and $10$ (right)
are shown. 
}
\label{3D_pic}
\end{figure}
%

Figure~\ref{3D_pic} shows the 3D plot of the generated data $g(z,\rho)$
for $z_0=1$ and $\alpha_{\rm GB}=0$, $1$, and $10$ in the case $D=5$. 
Here, the data of the ninth layer (layer of $n=9$) 
are used to draw this figure. As the
value of $\alpha_{\rm GB}$ is increased, the surface becomes
steeper around the puncture.

%
%
\subsection{Common apparent horizon}

The AH is defined as the outermost marginally
trapped surface, and it satisfies the equation 
of zero expansion, $\theta_+=\nabla_\mu k^\mu =0$, where $k^\mu$
is a tangent vector of the null geodesic congruence
from the AH. In GR, the formation of an AH
implies the existence of an event horizon (EH) outside of it
assuming the cosmic censorship and the null energy condition for
the energy-momentum tensor.
On the other hand, in GB gravity, this statement does not
hold because whether $-\mathcal{H}_{\mu\nu}$ obeys the null energy
condition is quite uncertain. However, also in GB gravity,
the formation of an AH at least implies the existence
of a region where gravity is strong. Furthermore, 
many theorems, such as the area theorem and the  
second law for a future trapping horizon, 
has been shown
to hold also in GB gravity
for a spherically-symmetric system with matter
of GR branch \cite{Nozawa:2007}.
For this reason, the AH may imply the black hole
formation also in GB in a certain condition.
For this reason, it is interesting to study the formation
of an AH also in GB gravity.

Because the space is momentarily static in our setup, 
the equation of an apparent 
horizon is reduced to $D_is^i=0$, where $s^i$ 
is a unit normal to the horizon. Assuming the
location of the AH to be $R=h(\theta)$,
where $\theta=\arctan(\rho/z)$, the horizon equation
becomes
\begin{multline}
h_{,\theta\theta}-
(D-2)(h^2+h_{,\theta}^2)
\left(\frac{2}{D-3}\frac{\Psi_{,R}}{\Psi}+\frac{1}{h}\right)
+\frac{h_{,\theta}^2}{h}\\
+h_{,\theta}\left(1+\frac{h_{,\theta}^2}{h^2}\right)
\left[\frac{2(D-2)}{(D-3)}\frac{\Psi_{,\theta}}{\Psi}
+(D-3)\cot\theta\right]=0.
\end{multline}
The common AH that encloses the two black holes
is found using the so-called shooting
method by solving this equation under the boundary condition
$h_{,\theta}=0$ at $\theta=0$ and $\pi/2$.

%
\begin{figure}[tb]
\centering
{
\includegraphics[width=0.35\textwidth]{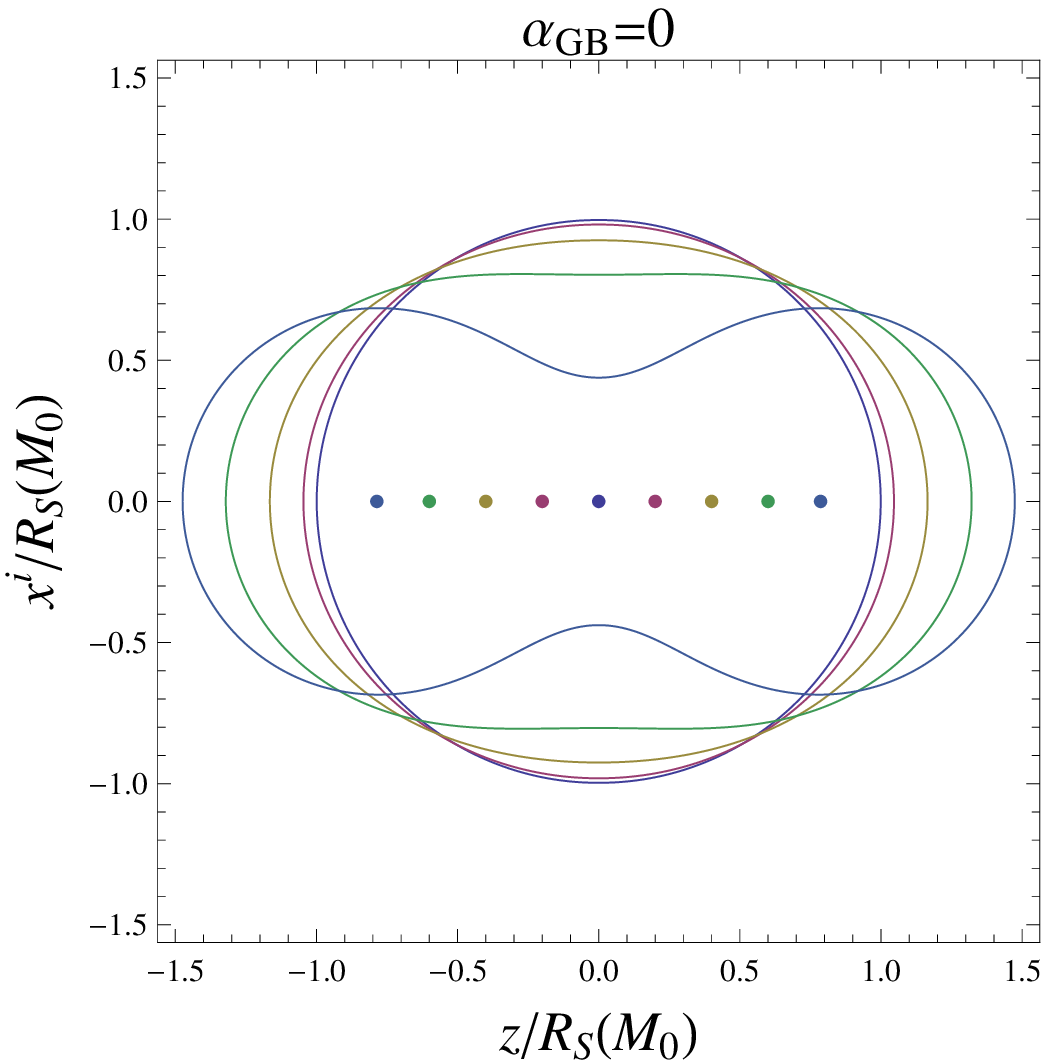}
\hspace{5mm}
\includegraphics[width=0.35\textwidth]{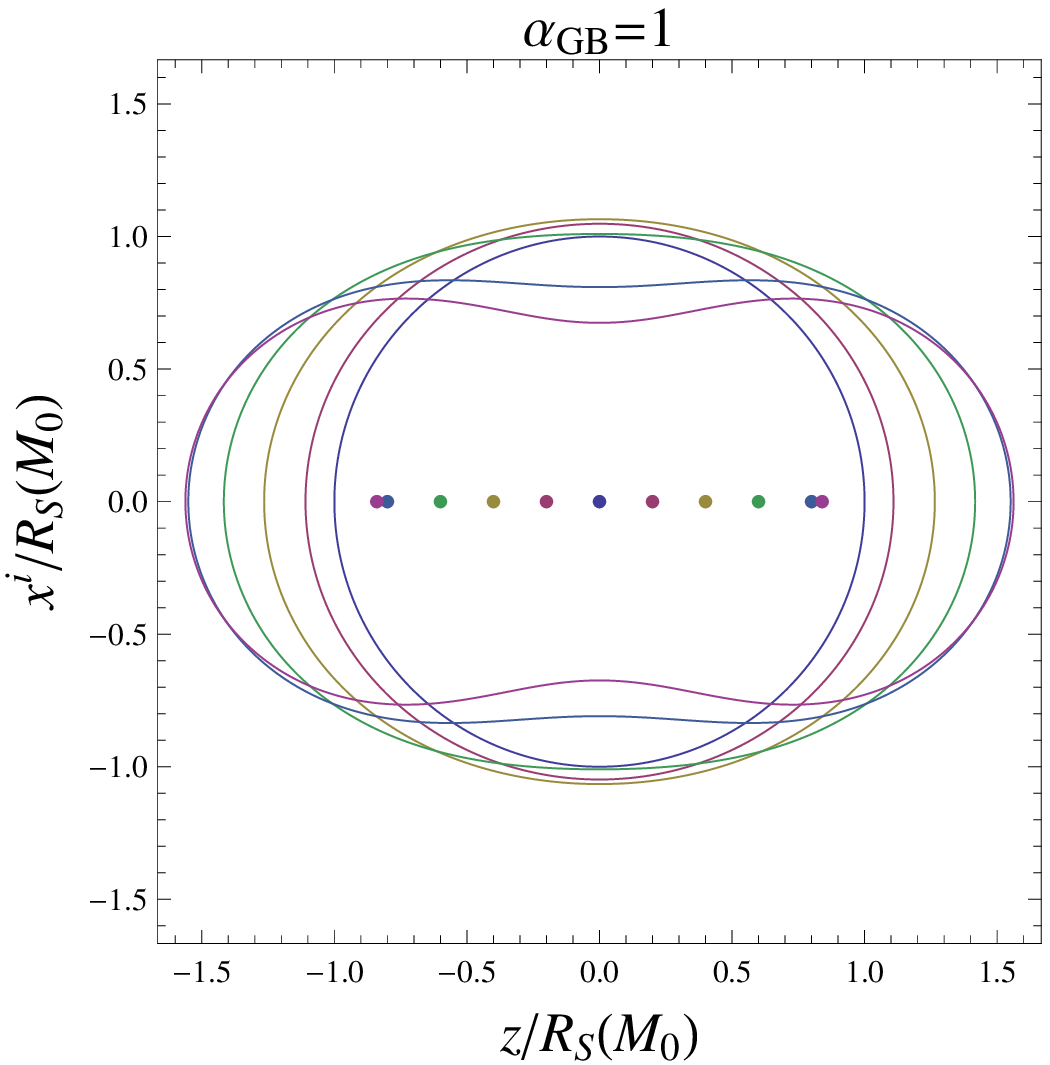}
\includegraphics[width=0.35\textwidth]{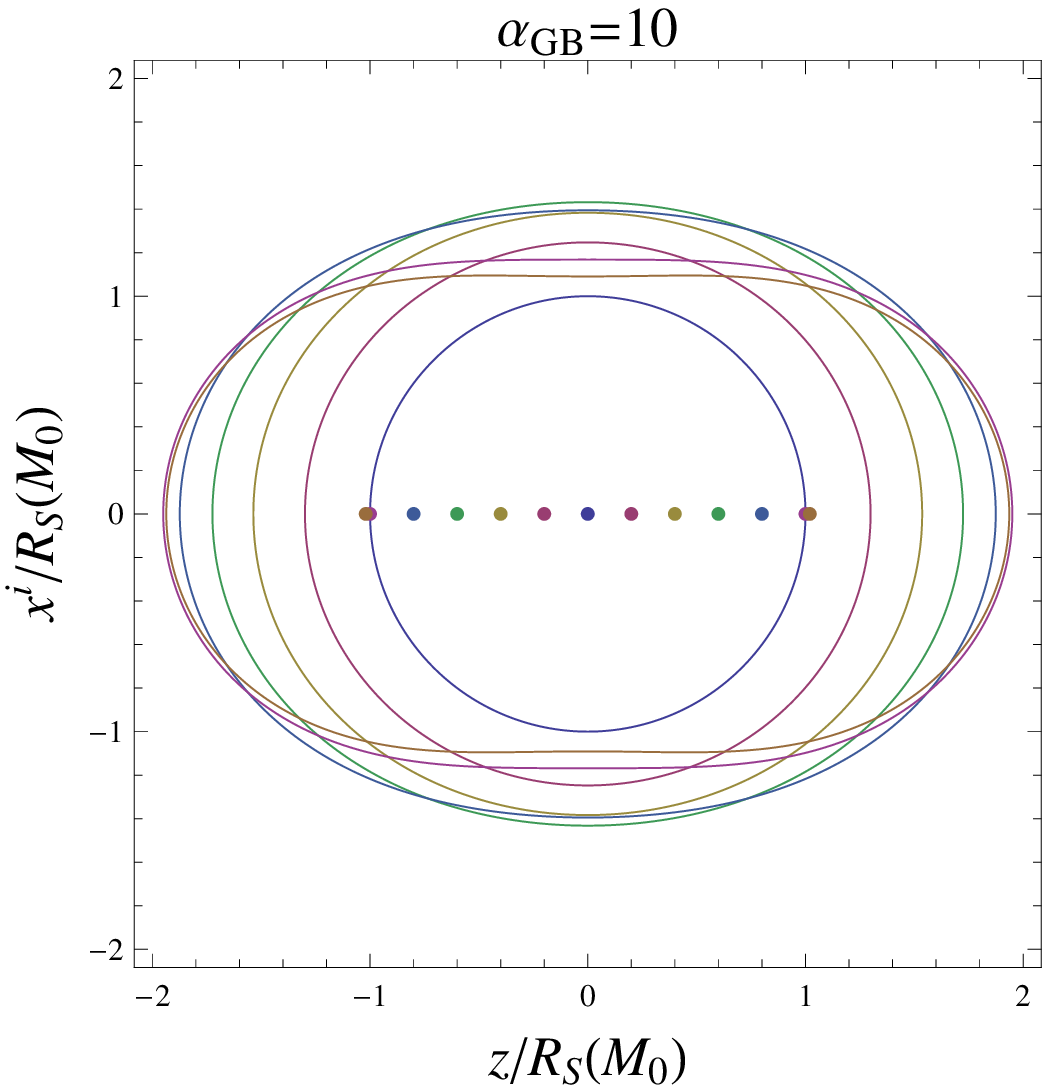}
\hspace{5mm}
\includegraphics[width=0.35\textwidth]{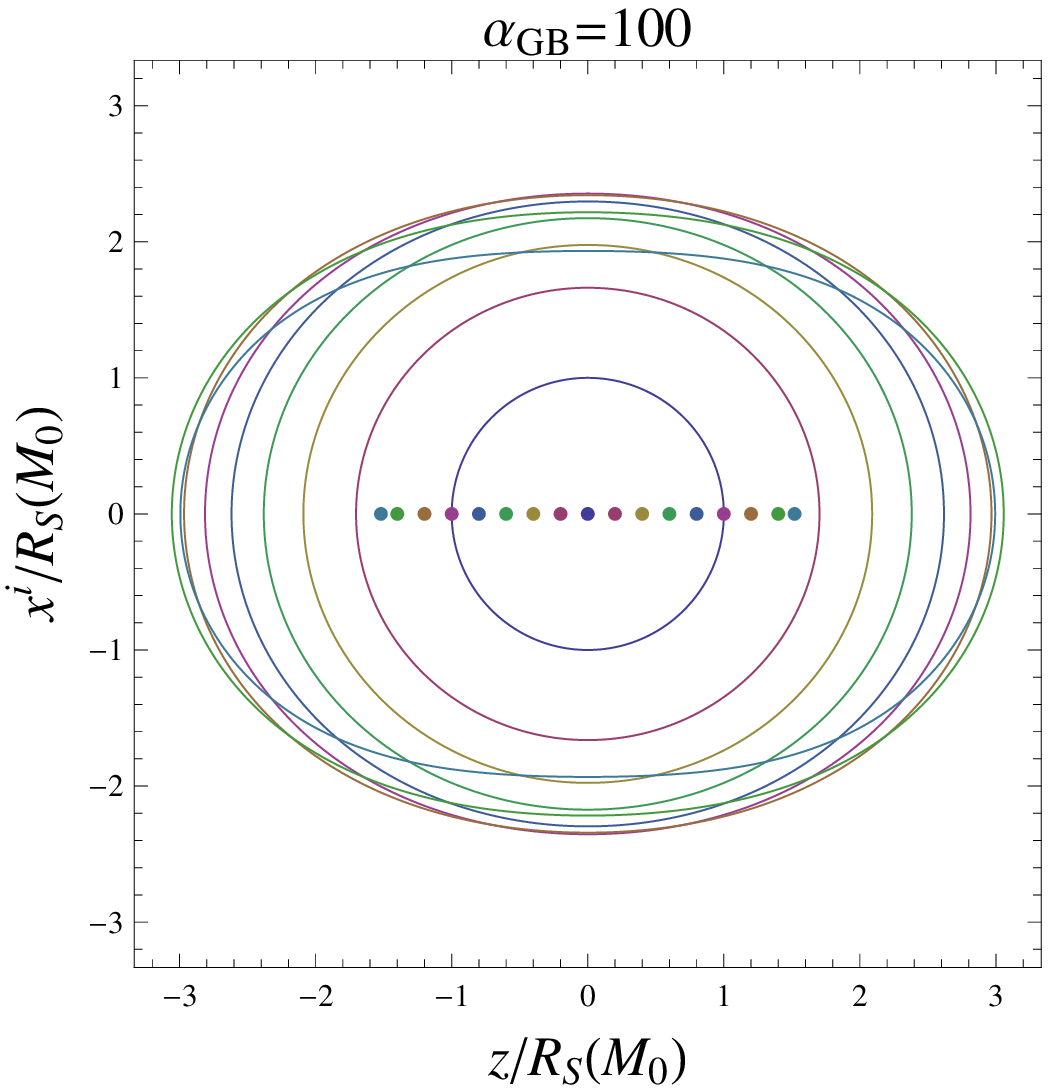}
}
\caption{Coordinate shapes of the AH on a $(z, x^i)$ plane in
the unit of $R_S(M_0)=1$ in the case of $D=5$ and
$\alpha_{\rm GB}=0$, $1$, $10$, and $100$. The cases for several values
of $z_0$ are shown with $0.2$ intervals starting from zero, and 
the case $z_0=z_0^{\rm (crit)}$ is also shown. The location
of the punctures are shown by dots. As $z_0$ is increased,
the horizon becomes distorted. }
\label{AH-shape}
\end{figure}
%

Figure~\ref{AH-shape} shows examples of coordinate shape 
of the common AH on the $(z,x^i)$ plane,
where $x^i$ is one of the orthogonal directions to the $z$ axis.
Here, the cases $D=5$ and $\alpha_{\rm GB}=0$, $1$, $10$,
and $100$ are shown, and the values of $z_0$ are $0$, $0.2$,..., and
$z_0^{\rm (crit)}$, where $z_0^{\rm (crit)}$
is the critical value  
for the AH formation.
As the value of $z_0$ is increased, 
the AH becomes more distorted.
There is the other solution
to the AH equation, which corresponds to
the inner boundary of the trapped region. 
At $z_0=z_0^{\rm (crit)}$, the two solutions degenerate
and the solution vanishes for $z_0>z_0^{\rm (crit)}$.

%
\begin{figure}[tb]
\centering
{
\includegraphics[width=0.35\textwidth]{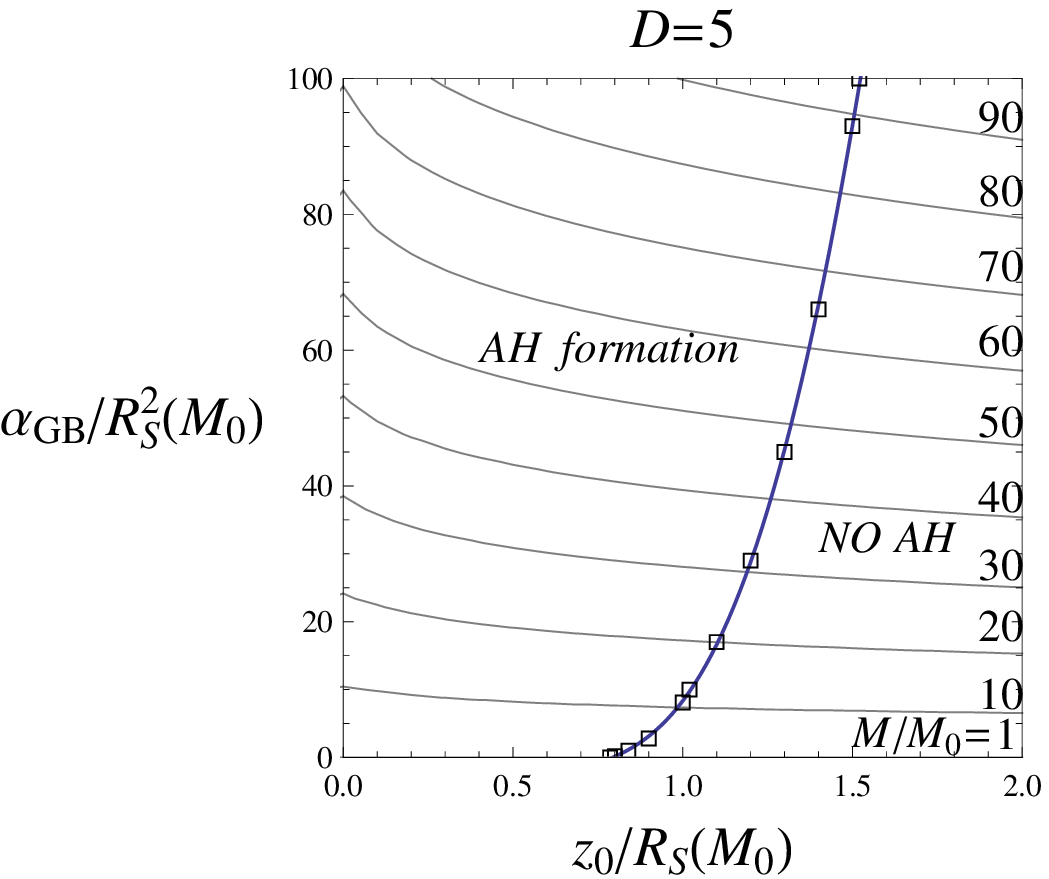}
\includegraphics[width=0.35\textwidth]{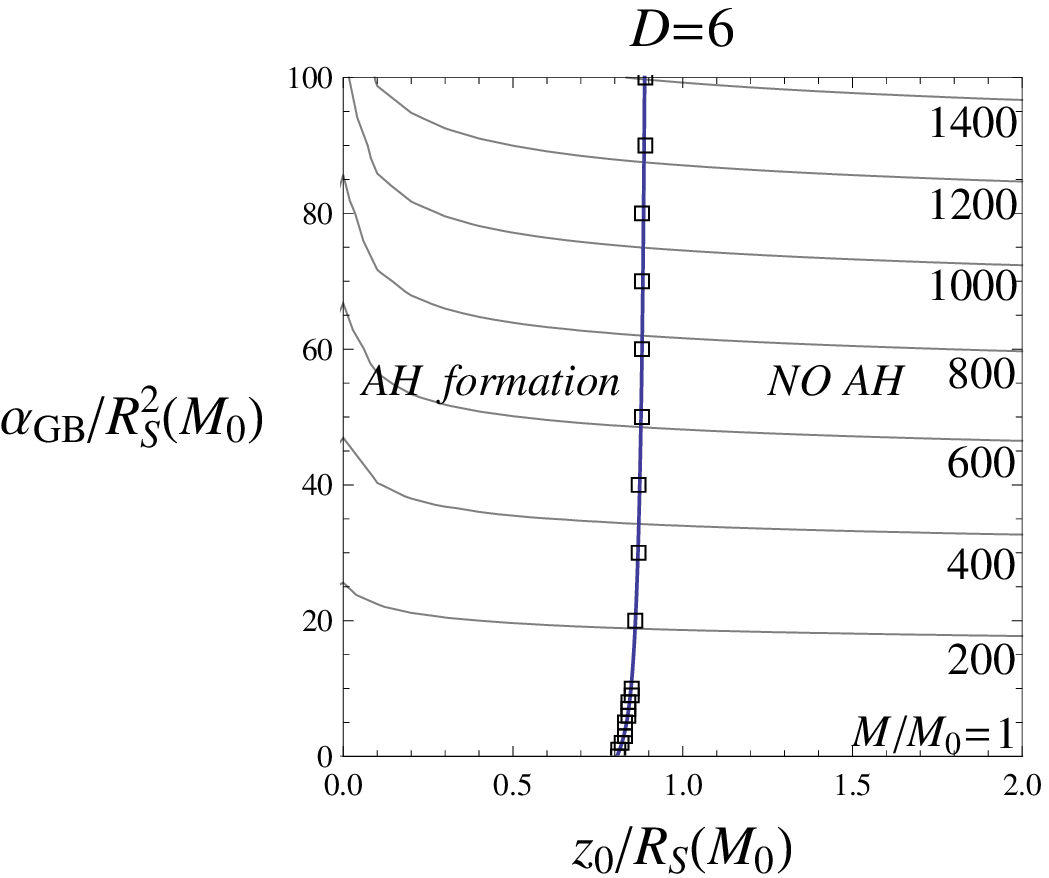}
\includegraphics[width=0.35\textwidth]{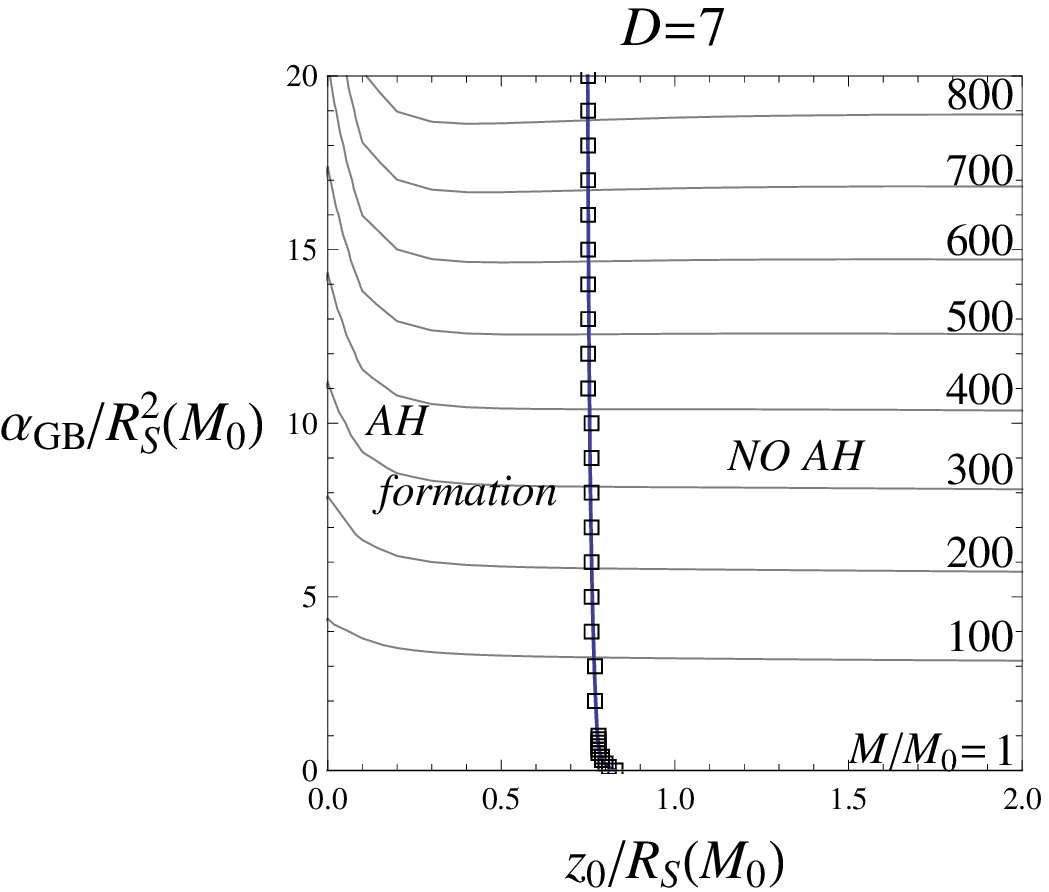}
\includegraphics[width=0.35\textwidth]{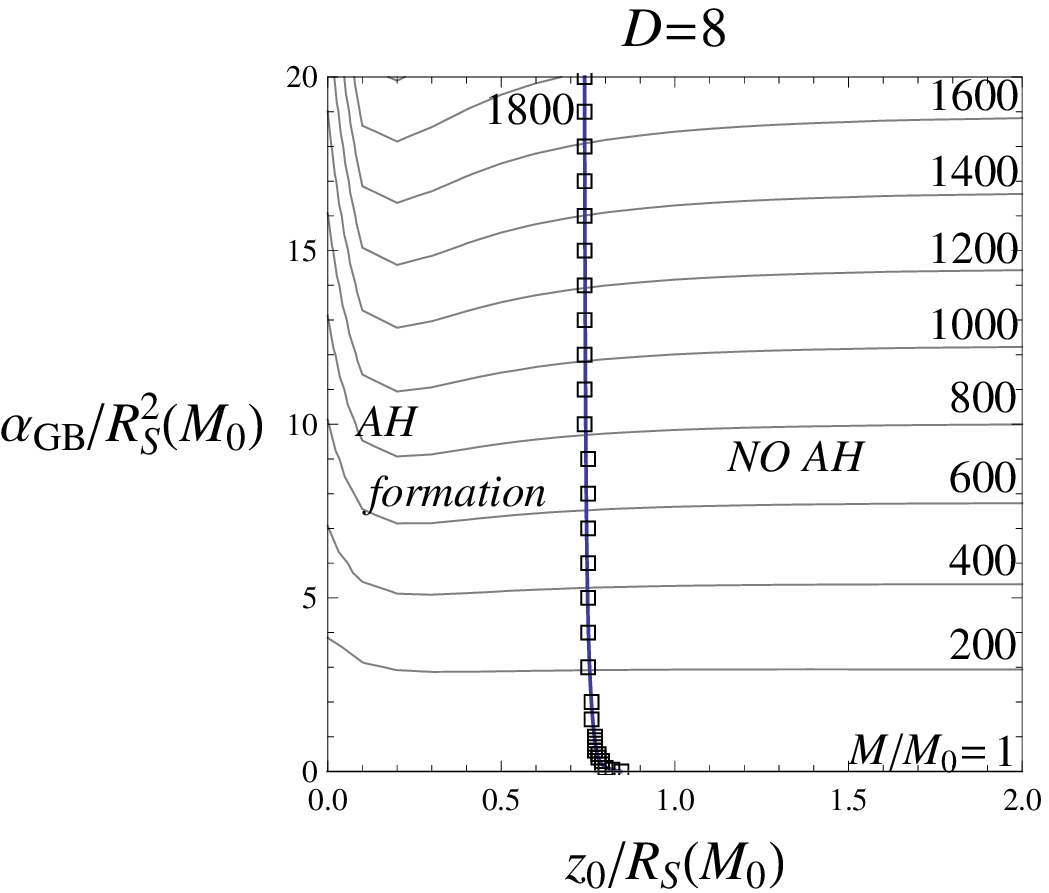}
}
\caption{The regions where the AH can be found (AH formation)
and cannot be found (NO AH) and contours of
$M/M_0$ on a $(z_0, \alpha_{\rm GB})$ plane in the unit $R_S(M_0)$. 
The AH cannot be found for $z_0>z_0^{\rm (crit)}$, and numerical data
of $z_0^{\rm (crit)}$ are shown by squares ($\square$). The ADM mass
becomes large as $\alpha_{\rm GB}$ is increased. 
}
\label{Z0_AGB_UnitM0}
\end{figure}
%

Figure~\ref{Z0_AGB_UnitM0} shows the
region where the AH can be found on the $(z_0,\alpha_{\rm GB})$ plane
for $D=5$--$8$.
Here, the unit of the length is adopted as $R_S(M_0)$.
In general dimensions, the AH is not formed for sufficiently
large $z_0$. In the cases $D=5$ and $6$, the value of
$z_0^{\rm (crit)}/R_S(M_0)$ becomes large
as $\alpha_{\rm GB}$ is increased. At first glance,
one may think that increasing the coupling constant
$\alpha_{\rm GB}$ helps the AH formation. However,
this interpretation is not correct, because 
in this figure, the artificial mass $M_0$ is used
in the length unit $R_S(M_0)$. As we have seen in Sec.~\ref{Sec:ADM},
the ADM mass $M$ is changed as $\alpha_{\rm GB}$ is increased
following Eq.~\eqref{ADM_mass}. Several contours of 
$M/M_0$ are shown in Fig.~\ref{Z0_AGB_UnitM0}. 
As the value of $\alpha_{\rm GB}$ is increased, the 
mass $M$ also becomes large.

%
\begin{figure}[tb]
\centering
{
\includegraphics[width=0.35\textwidth]{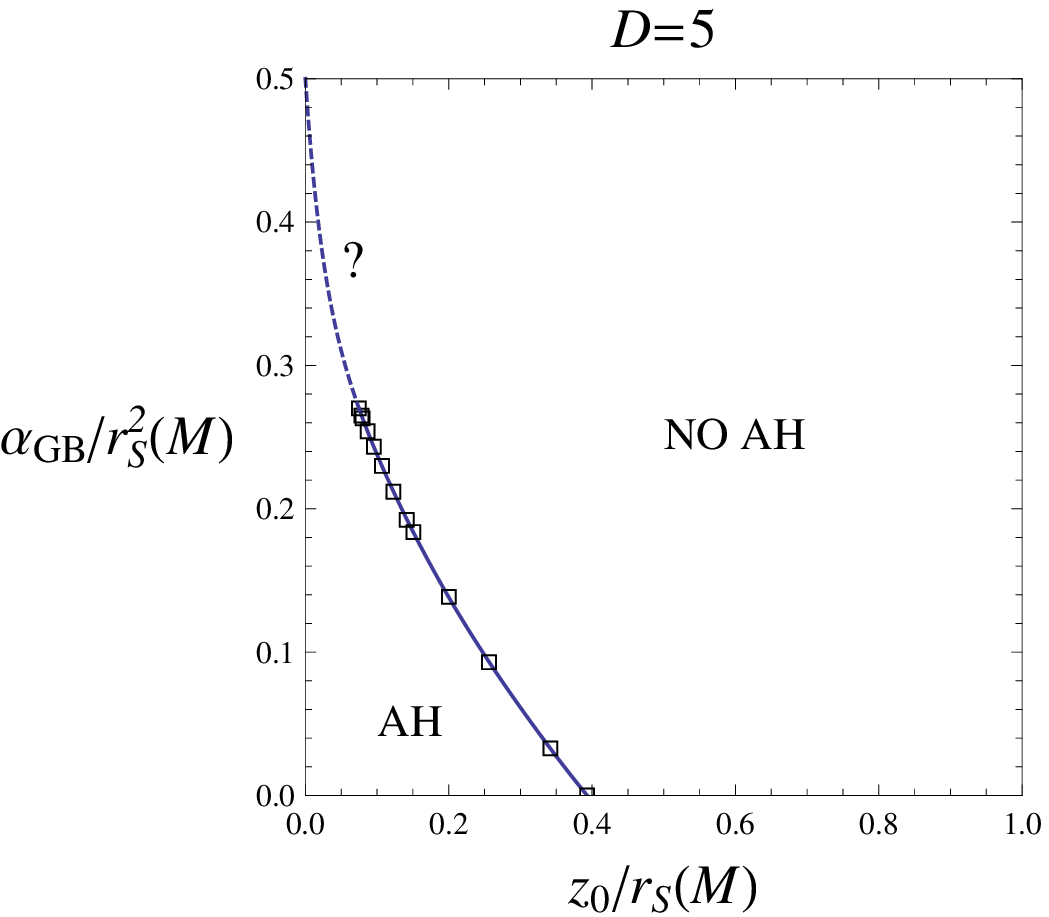}
\includegraphics[width=0.35\textwidth]{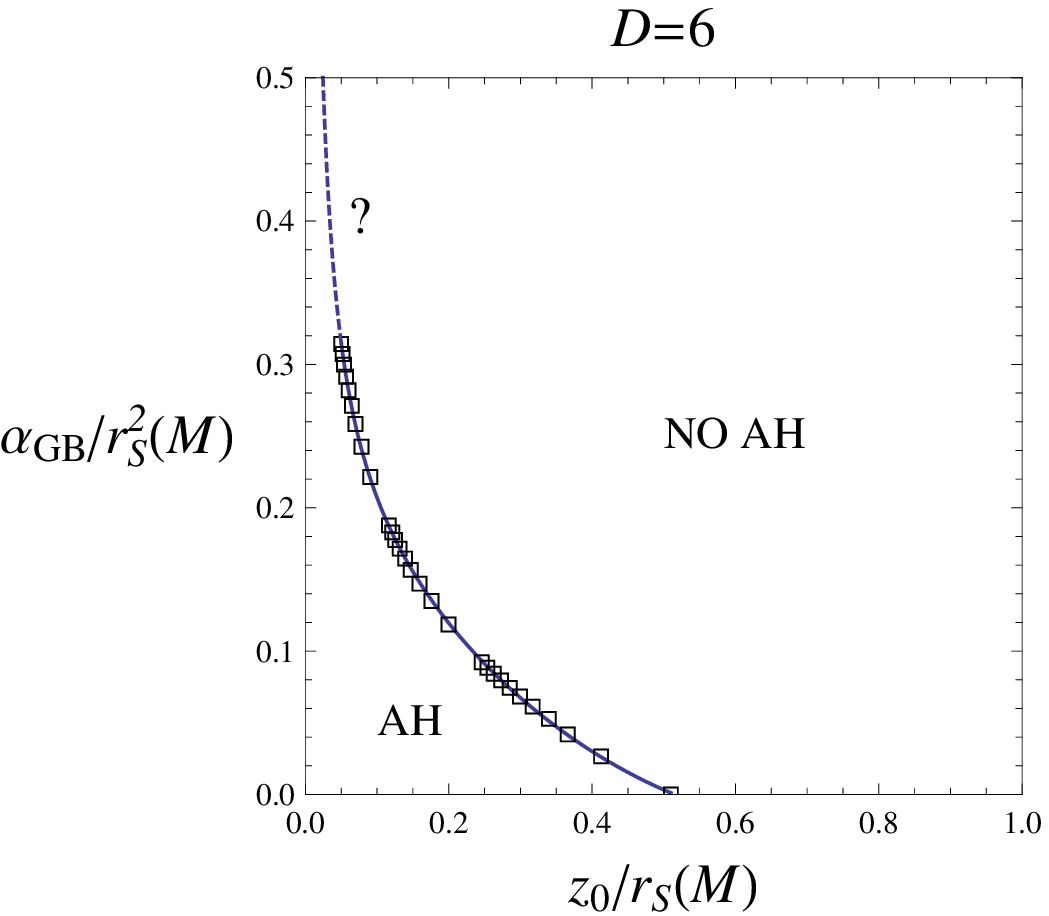}
\includegraphics[width=0.35\textwidth]{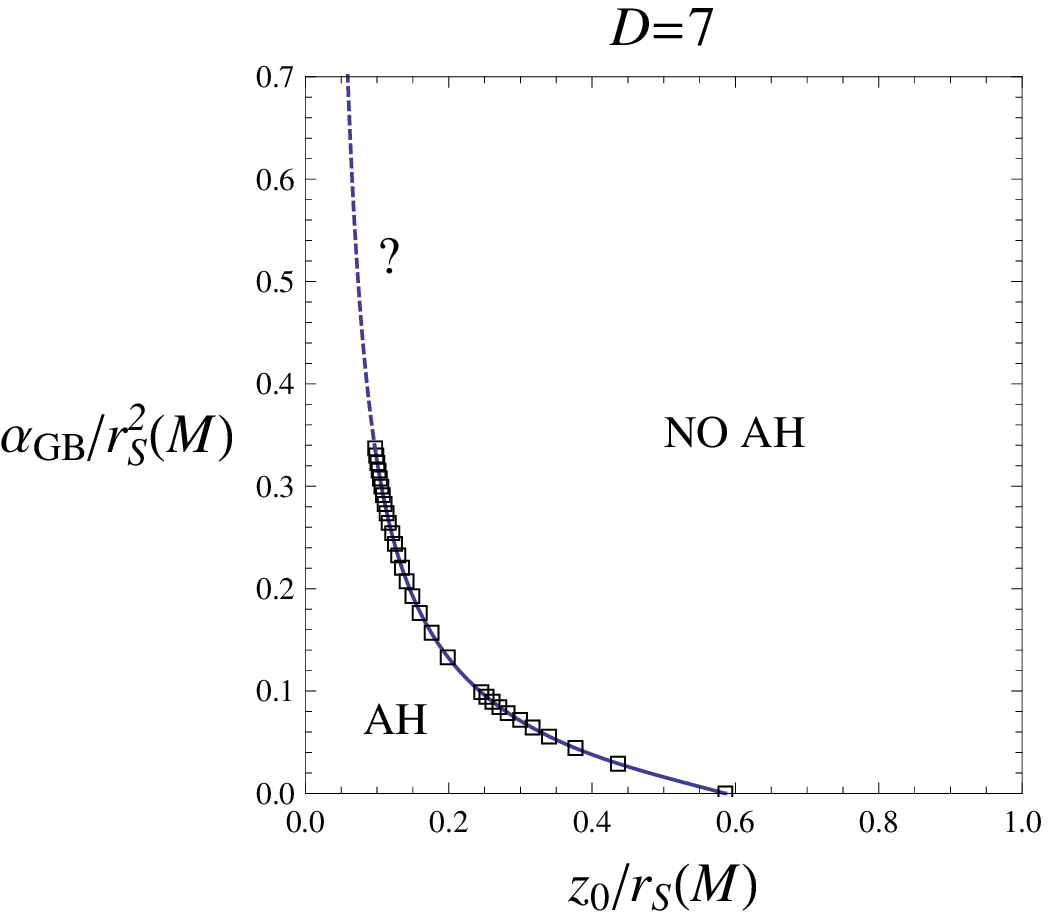}
\includegraphics[width=0.35\textwidth]{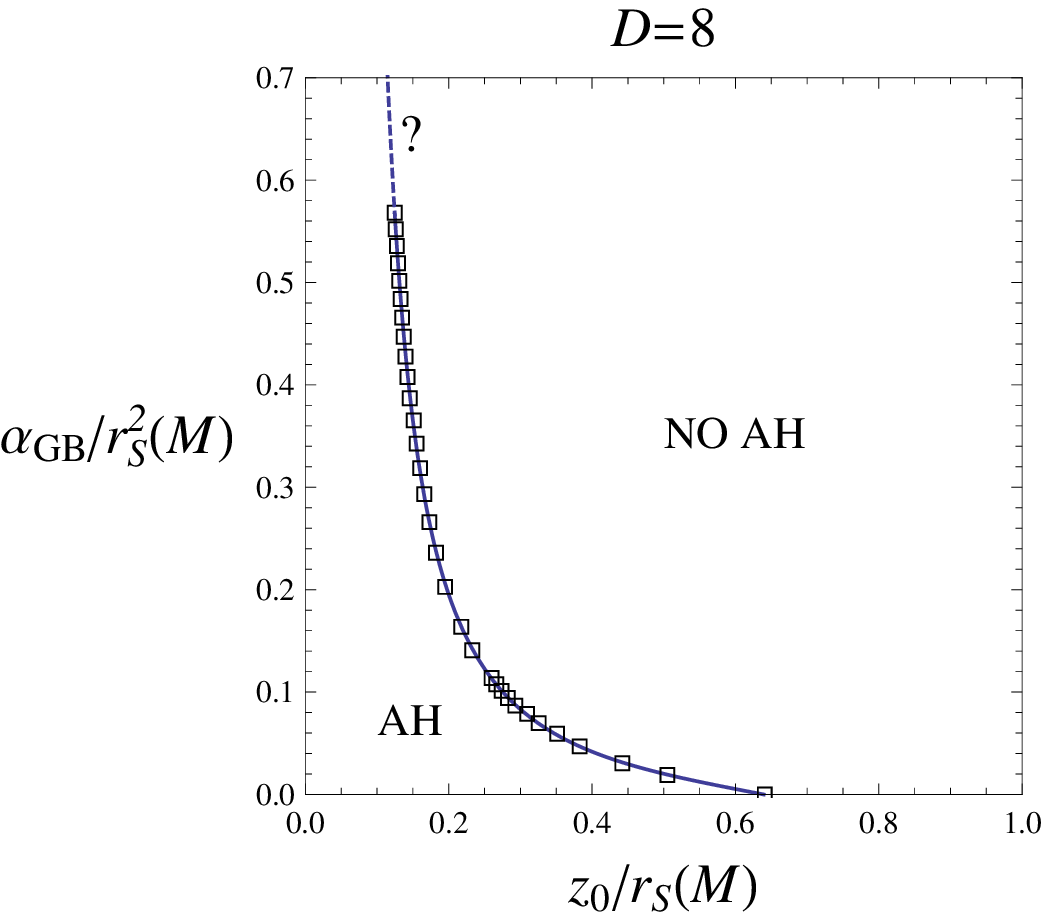}
}
\caption{The regions where the AH can be found (AH)
and cannot be found (NO AH) 
on a $(z_0, \alpha_{\rm GB})$ plane in the unit $r_S(M)$.
The numerical data are shown by squares  ($\square$), and solid line shows the
border line of the region of AH formation drawn by interpolation. 
The dotted line is expected border line for large $\alpha_{\rm GB}$
for which numerical data were not taken. The border line is
expected to intersect
$z_0$ axis at $\alpha_{\rm GB}/r_S^2(M)=0.5$ in the case $D=5$, 
and not to cross the $z_0$ axis for $D\ge 6$.}
\label{Z0_AGB_UnitADM}
\end{figure}
%

Figure~\ref{Z0_AGB_UnitADM} shows the region of
AH formation on the $(z_0, \alpha_{\rm GB})$ plane,
but now the Schwarzschild radius of the ADM mass, $r_S(M)$,
is used as the unit of the length. 
In all dimensions $D=5$--$8$, the value
$z_0^{\rm (crit)}$ decreases as the value of $\alpha_{\rm GB}$
is increased, indicating that the coupling constant $\alpha_{\rm GB}$
makes the AH formation difficult compared to the GR case.  
This is a natural result, because 
in the spherically symmetric case, the ratio of 
the horizon radius to the Schwarzschild radius, $r_H(M)/r_S(M)$,
decreases as $\alpha_{\rm GB}/r_S^2(M)$ is increased.
The border of the region of the AH formation is shown by
a solid curve by interpolating the numerical data.
The dashed line shows the expected border 
for large $\alpha_{\rm GB}$ for which
numerical calculation has not been done in this paper.
In the case $D=5$, it is naturally expected 
that the border line crosses $z_0$ axis
at $\alpha_{\rm GB}/r_S^2(M)=0.5$, because
for this value, the horizon radius of 
the spherically-symmetric black hole becomes zero.
For the other dimensions, the border line would not cross
$z_0$ axis, since $r_H(M)>0$ for arbitrary value of $\alpha_{\rm GB}$ 
(see Fig.~\ref{alpha_radius}).

%
%
\subsection{Penrose inequalities}

The Penrose inequality in GR,  
\begin{equation}
A_{\rm AH}\le \Omega_{D-2}\left[r_S(M)\right]^{D-2},
\label{Eq:penrose1}
\end{equation}
conjectures that the AH area is bounded from above by the
horizon area of a spherically-symmetric 
black hole (Schwarzschild-Tangherlini black hole)
with the same mass. The reason for this conjecture as follows.
If the cosmic censorship holds, an AH 
is formed in an EH, and since the EH
is located outside of the AH, its area is expected to be
larger than that of the AH. Because of the area theorem
by Hawking, the area of the EH will increase and asymptote
to that of a final stationary black hole. Since the
horizon area of a rotating black hole is smaller than that
of a Schwarzschild black hole with the same mass, and 
the final mass is smaller than the ADM mass because of the
gravitational radiation, the inequality \eqref{Eq:penrose1}
is expected to hold. On the other hand, if the system
with an AH whose area is greater than the bound of \eqref{Eq:penrose1},
the validity of the cosmic censorship, 
one of the assumptions of the above discussion,
may be suspected.

The Penrose inequality has been attracting a lot of attentions,
and it was proved for momentarily static initial data
for $D=4$--$7$ \cite{Bray:2007}.
On the other hand, a ``counterexample'' (but not in a strict sense)
has been found in Ref.~\cite{BenDov:2004} in a system consisting of combined
portions of the Schwarzschild and Oppenheimer-Snyder 
spacetimes. This example does not contradict the cosmic censorship,
but contradicts the assumption that the area of the EH 
is larger than that of the AH. But the Penrose inequality
can be modified to match this counterexample as follows.
The AH in a usual sense can be regarded as a ``black hole AH''
since it is formed in a black hole.
On the other hand, we can consider a ``white hole AH''
formed in a white hole 
whose past-directed outgoing null geodesic congruence
has zero expansion. In the above counterexample, 
the black hole AH is located inside of the white hole AH,
and the area of the white hole AH satisfies the Penrose inequality.
Therefore, if we adopt the outermost horizon, the Penrose inequality
still holds. This Penrose inequality for the outermost horizon
can be proved in spherically-symmetric case assuming
weak energy condition \cite{Malec:1994}.

In GB gravity, the relation between the Penrose inequality
and the cosmic censorship becomes unclear because the
above reasoning for the Penrose inequality 
may not hold since $-\mathcal{H}_{\mu\nu}$ may violate
the energy condition. 
However, it is still of interest whether the universal
relation like \eqref{Eq:penrose1} holds 
or not from a mathematical point of view.
The proof for the GR case cannot be 
applied to the GB case at least straightforwardly, because it is unclear
if $-\mathcal{H}_{\mu\nu}$ satisfies the energy condition.
Therefore, it is interesting to test the inequality using
the initial data constructed in this paper.

Other than the original version of the Penrose inequality 
\eqref{Eq:penrose1}, 
we can consider another inequality that 
may be expected to hold in GB 
gravity.\footnote{The author thanks Tetsuya Shiromizu for this point.} 
Namely, since the horizon radius $r_H(M)$ of a spherically-symmetric
black hole is different from $r_S(M)$, the Penrose inequality
may be modified as
\begin{equation}
A_{\rm AH}\le \Omega_{D-2}\left[r_H(M)\right]^{D-2},
\label{Eq:penrose2}
\end{equation}
where $r_H(M)$ is defined by Eq.~\eqref{Eq:horizon_radius}.
In the following, we test if these two inequalities \eqref{Eq:penrose1}
and \eqref{Eq:penrose2} hold in our system.
For this purpose,
we define $P_1:=A_{\rm AH}/\Omega_{D-2}\left[r_S(M)\right]^{D-2}$
and $P_2:=A_{\rm AH}/\Omega_{D-2}\left[r_H(M)\right]^{D-2}$
and evaluate these values for selected values of $z_0$
and $\alpha_{\rm GB}$.

%
\begin{figure}[tb]
\centering
{
\includegraphics[width=0.4\textwidth]{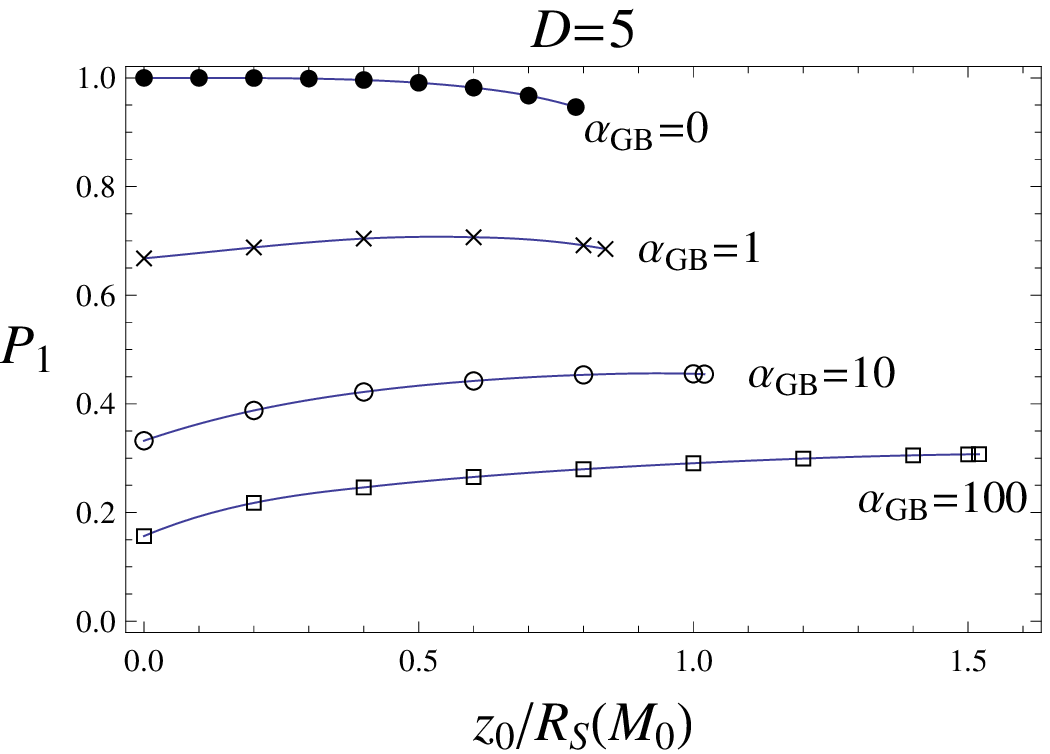}
\hspace{5mm}
\includegraphics[width=0.4\textwidth]{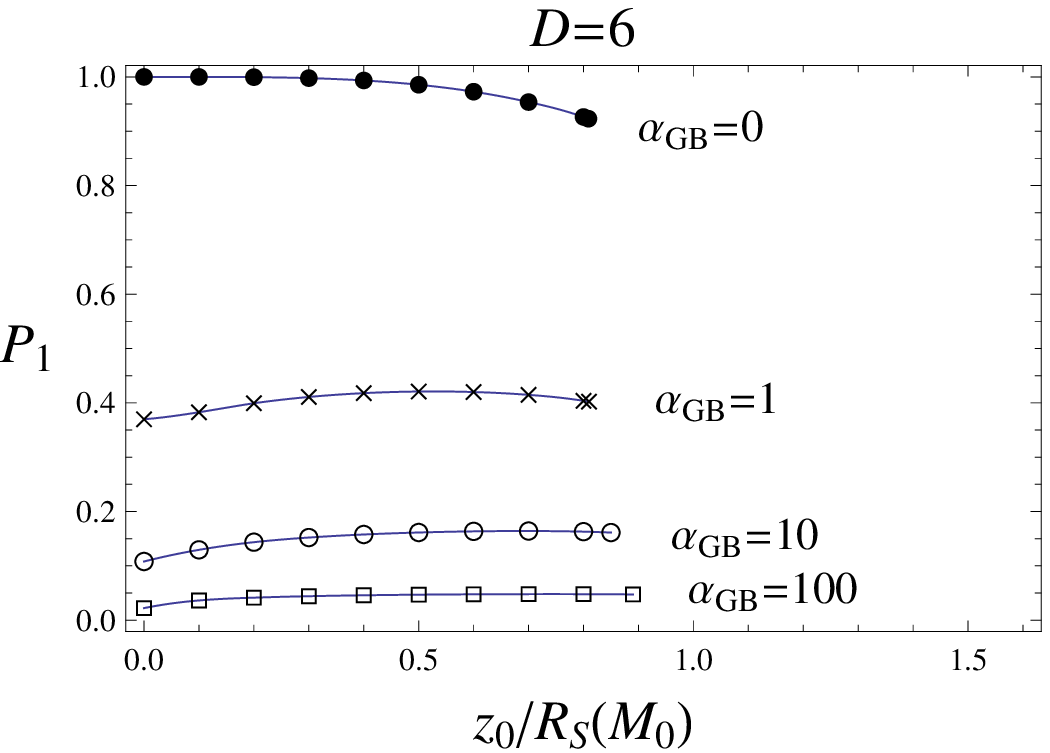}
\includegraphics[width=0.4\textwidth]{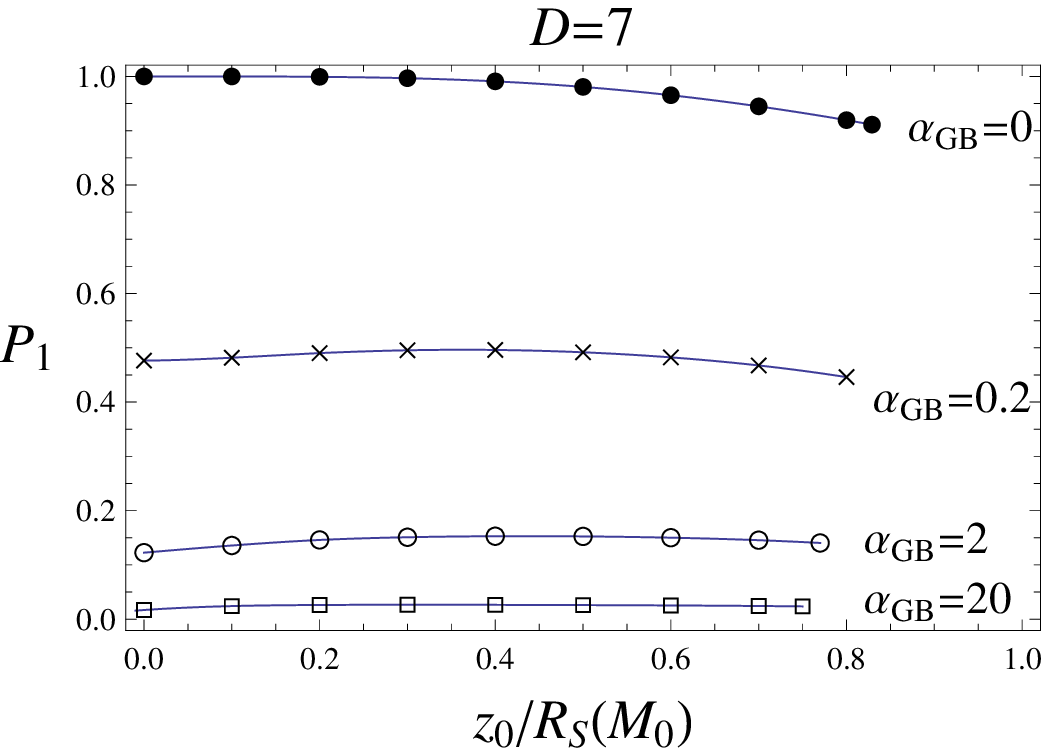}
\hspace{5mm}
\includegraphics[width=0.4\textwidth]{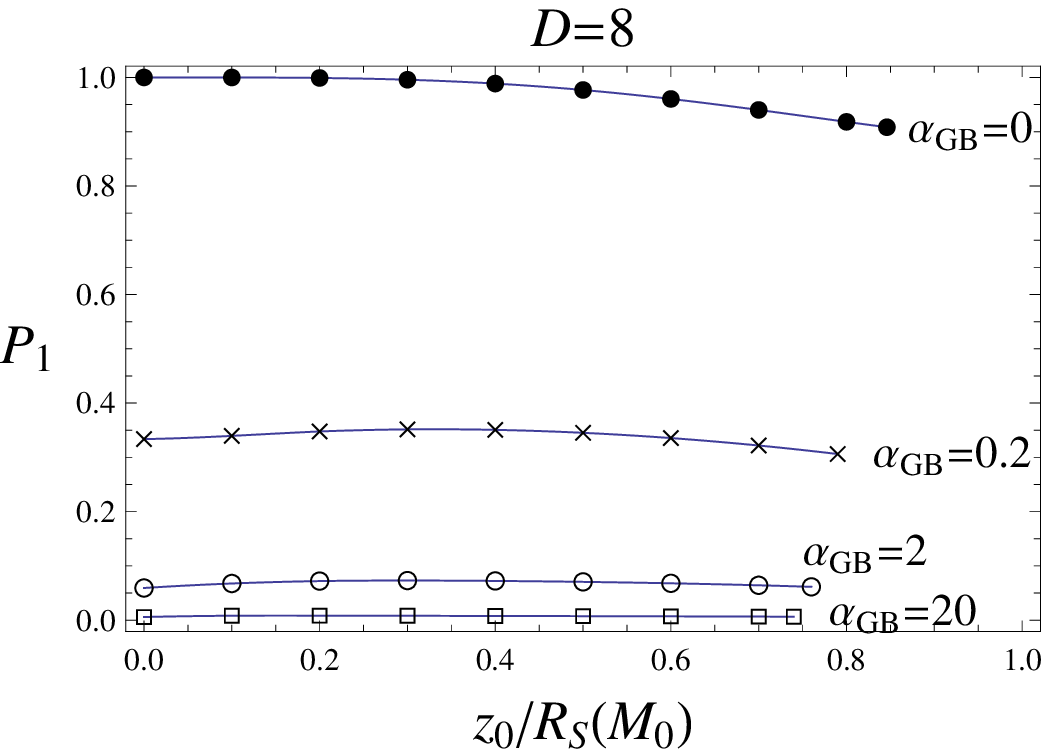}
}
\caption{The behavior of $P_1:=A_{\rm AH}/\Omega_{D-2}[r_S(M)]^{D-2}$
as a function of $z_0$ for $\alpha_{\rm GB}=0$ ($\bullet$), $1$
($\times$), $10$ ($\circ$), and $100$ ($\square$)
in the cases $D=5$ and $6$ and for
$\alpha_{\rm GB}=0$ ($\bullet$), $0.2$ ($\times$), $2$ ($\circ$), 
and $20$ ($\square$)
in the cases $D=7$ and $8$. 
The value of $P_1$ is not greater than unity in all cases,
suggesting that the Penrose inequality $P_1\le 1$ holds
in this system. 
}
\label{penrose1}
\end{figure}
%

Figure~\ref{penrose1} shows the behavior of $P_1$
as a function of $z_0$. Here the cases of $\alpha_{\rm GB}=0$, $1$, $10$, and
$100$ are shown for $D=5$ and $6$, and
the cases of $\alpha_{\rm GB}=0$, $0.2$, $2$, $20$
are shown for $D=7$ and $8$. In all cases, the values of $P_1$
are smaller than unity, suggesting that
the inequality~\eqref{Eq:penrose1} is kept in this system.
We find that $P_1$ becomes smaller 
as $\alpha_{\rm GB}$ is increased, and therefore,
the coupling constant tends to help 
the AH satisfy the inequality \eqref{Eq:penrose1} 
if $\alpha_{\rm GB}$ is positive.

%
\begin{figure}[tb]
\centering
{
\includegraphics[width=0.4\textwidth]{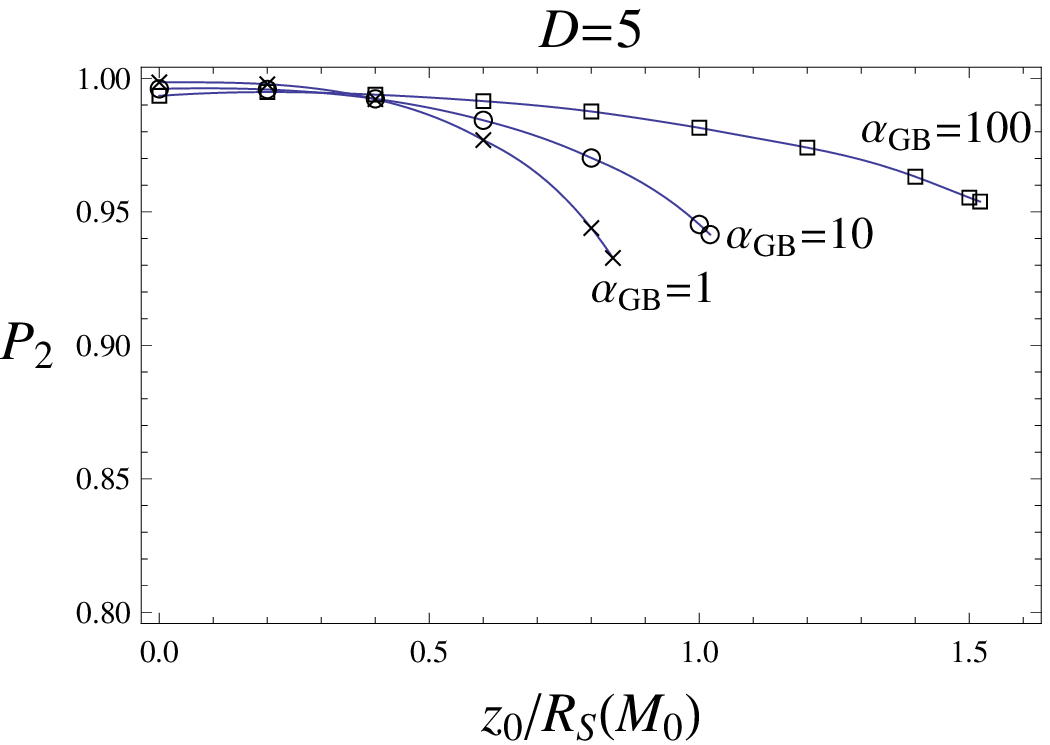}
\hspace{5mm}
\includegraphics[width=0.4\textwidth]{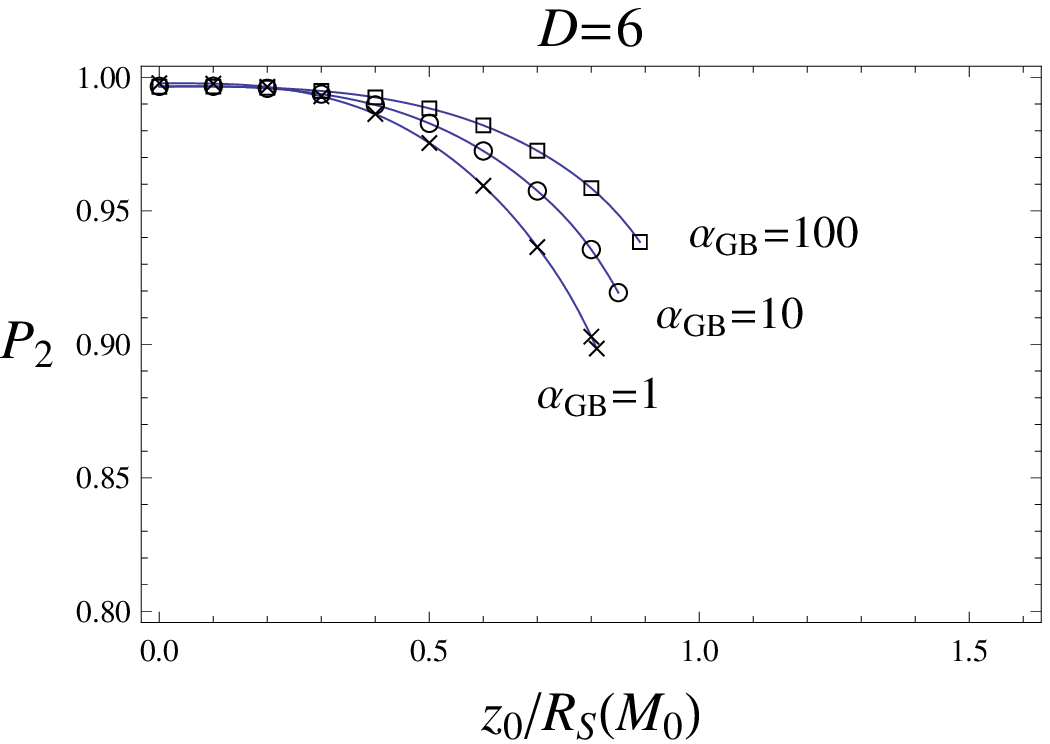}
\includegraphics[width=0.4\textwidth]{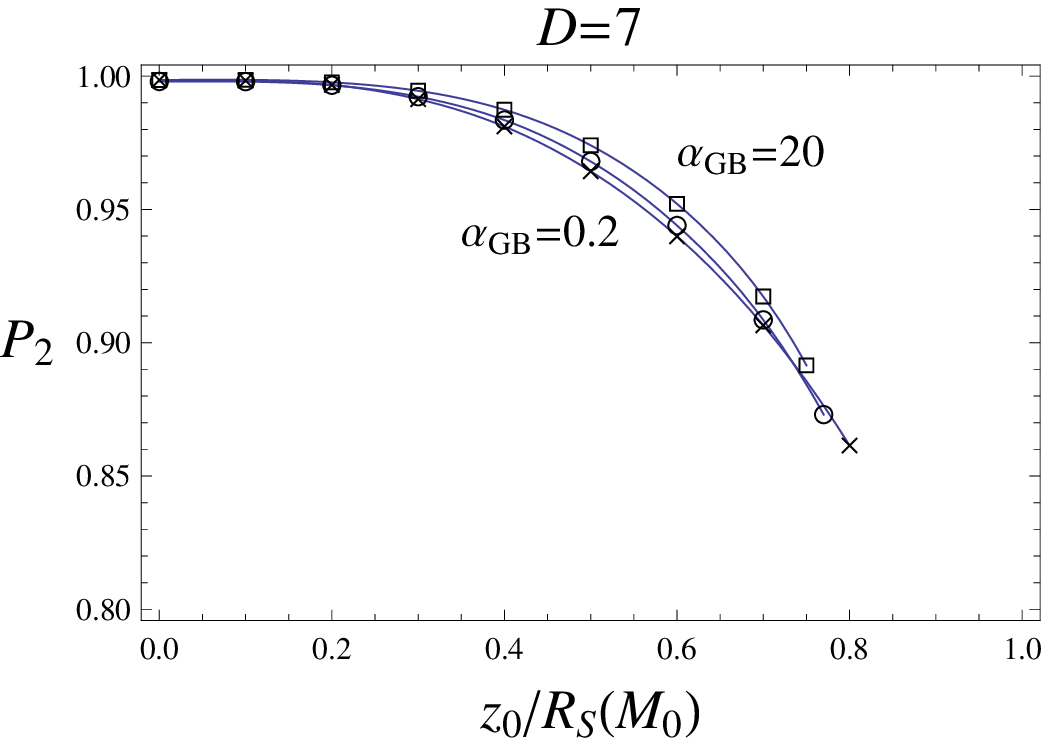}
\hspace{5mm}
\includegraphics[width=0.4\textwidth]{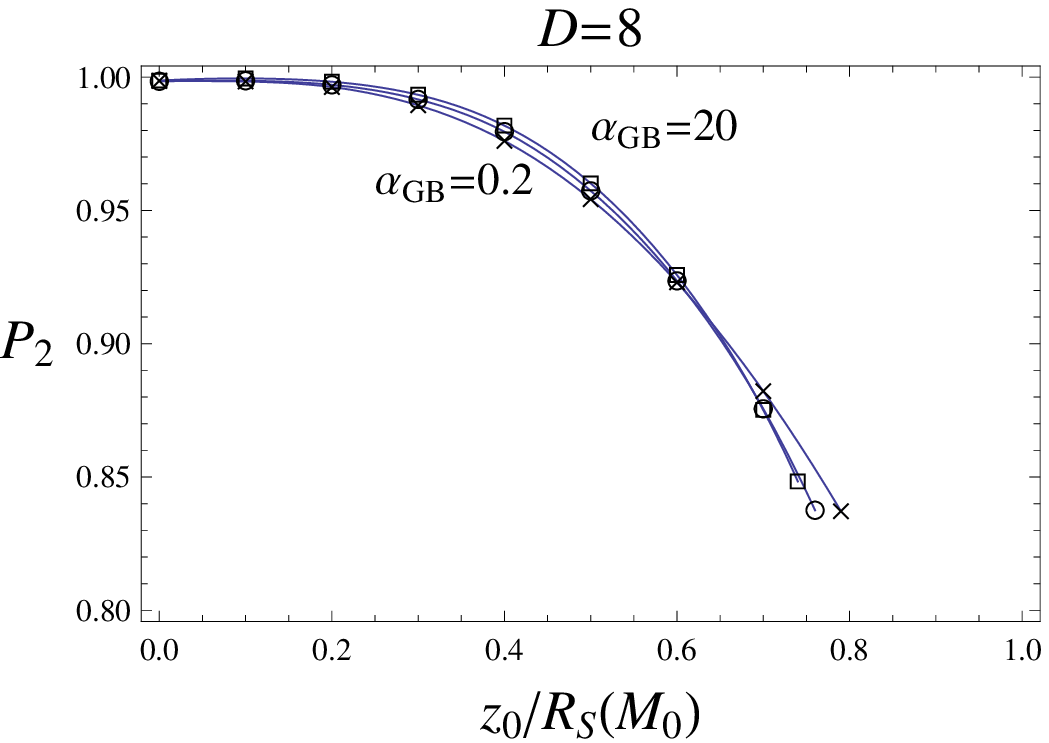}
}
\caption{The behavior of $P_2:=A_{\rm AH}/\Omega_{D-2}[r_H(M)]^{D-2}$
as a function of $z_0$ for $\alpha_{\rm GB}=1$ ($\times$), $10$
($\circ$), and $100$ ($\square$)
in the cases $D=5$ and $6$ and for
$\alpha_{\rm GB}=0.2$ ($\times$), $2$ ($\circ$), and $20$ ($\square$)
in the cases $D=7$ and $8$. 
The value of $P_2$ is unity for $z_0=0$ and decreases
as $z_0$ is increased. Therefore
the modified version of the Penrose inequality $P_2\le 1$ 
also holds in this system. }
\label{penrose2}
\end{figure}
%

Figure~\ref{penrose2} shows the behavior of $P_1$
as a function of $z_0$. Here the cases of $\alpha_{\rm GB}=1$, $10$, and
$100$ are shown for $D=5$ and $6$, and
the cases of $\alpha_{\rm GB}=0.2$, $2$, $20$
are shown for $D=7$ and $8$. 
For all $D$, the value of $P_2$
is unity for $z_0=0$ because the space agrees with the time-symmetric
slice of the spherically
symmetric black hole spacetime. As $z_0$ is increased,
the value of $P_2$ becomes smaller for all values of $\alpha$.
Therefore, the distortion of the AH makes the AH area smaller,
and the modified version of the Penrose inequality 
also holds in our system.

To summarize, the black hole initial data in this paper
satisfy both two Penrose inequalities \eqref{Eq:penrose1}
and \eqref{Eq:penrose2}, and no counterexample has been 
detected. Therefore, the Penrose inequalities might hold
also in GB gravity under appropriate conditions.

%
%
\section{Summary and discussion}

In this paper, we studied the method for generating the
initial data for one-black-hole and 
two-black-hole systems in GB gravity. 
Assuming the initial space to be momentarily static
and conformally flat, the highly nonlinear equation 
of the Hamiltonian constraint in the $N+1$ formalism was
successfully solved numerically. Using the generated
initial data, we studied the common AH 
that encloses the two black holes, and discussed
the Penrose inequalities in GB gravity.
The result suggests that both two inequalities
\eqref{Eq:penrose1} and \eqref{Eq:penrose2}
hold in this system.

Here, let us discuss whether the proposed conjecture
for the condition
of the AH formation in GR can be generalized to 
GB gravity. 
In four-dimensional GR, the hoop conjecture \cite{Thorne:72} is well known
as the condition of the horizon formation. The hoop
conjecture states that a black hole with horizon forms
when and only when the hoop length $C$ for a system
satisfies $C\lesssim 2\pi r_S(M)$. This conjecture
is loosely formulated in the sense that the definitions
of the horizon (AH or EH), the hoop, and the mass are not
explicitly specified. But this conjecture is known to 
give the approximate condition of the AH formation
(under appropriate definitions of the mass and the hoop which
may depend on researchers), although not explicitly proved.
The point in this conjecture is that the typical one-dimensional 
length of the system is restricted from above if an AH
is formed, implying that an arbitrarily long AH
does not form. But this statement holds only for $D=4$, because
the black hole can be arbitrarily long in higher dimensions
as expected from the black string solution, and explicitly
shown in \cite{IN02}. Instead,
Ida and Nakao proposed the generalized hoop conjecture \cite{IN02}
as 
\begin{equation}
C_{D-3}\lesssim \Omega_{D-3}[r_S(M)]^{D-3},
\label{hyperhoop}
\end{equation}
where $C_{D-3}$ is the typical $(D-3)$-dimensional 
quantity (``hyperhoop'') in this system.
This hyperhoop conjecture was discussed in several papers 
\cite{YN04,IN02,YN03,Yoo:2005,Yamada:2009},
and the results support its effectiveness.
Let us consider a two-black-hole system with total mass $M$, and 
suppose the two black holes to be momentarily at rest 
with a typical distance $L$. In four dimensions, the typical hoop length
is estimated by $C\simeq 2\pi r_S(M/2)+2L$. In a similar manner,
in $D$ dimensions, 
the typical hyperhoop quantity would be 
$C_{D-3}\simeq \Omega_{D-3}[r_S(M/2)]^{D-3}
+\Omega_{D-4}[r_S(M/2)]^{D-4}L$.
Substituting this formula into Eq.~\eqref{hyperhoop}, one has
\begin{equation}
L\lesssim \frac{\Omega_{D-3}}{2^{1/(D-3)}\Omega_{D-4}}r_S(M) \sim r_S(M).
\end{equation}
This gives an approximate condition for the AH formation
at least in a qualitative sense.

Can the hyperhoop conjecture be further extended
to GB gravity? One simple manner of generalization would be
to change from $r_S$ to $r_H$ in Eq.~\eqref{hyperhoop} as
\begin{equation}
C_{D-3}\lesssim \Omega_{D-3}[r_H(M)]^{D-3},
\label{GB-hyperhoop}
\end{equation}
where $r_H(M)$ is defined in Eq.~\eqref{Eq:horizon_radius}.
However, this simple generalization does not work
for a two-black-hole system.
Since the value of $C_{D-3}$ in this case is given by
$C_{D-3}\simeq \Omega_{D-3}[r_H(M/2)]^{D-3}
+\Omega_{D-4}[r_H(M/2)]^{D-4}L$, the condition 
\eqref{GB-hyperhoop} gives
\begin{equation}
L\lesssim \left(\frac{\Omega_{D-3}}{\Omega_{D-4}}\right)
\frac{[r_H(M)]^{D-3}-[r_H(M/2)]^{D-3}}{[r_H(M/2)]^{D-4}}.
\label{GB_condition_L}
\end{equation}
Let us consider the case $D=5$ and suppose $\alpha_{\rm GB}$
and $M$ satisfy the relation $\alpha_{\rm GB}=(1/2)r_S^{2}(M/2)$.
In this case, $r_H(M/2)=0$ and $r_H(M)=r_S(M)/\sqrt{2}$, and 
Eq.~\eqref{GB_condition_L} gives $L\lesssim \infty$. 
Therefore, the condition \eqref{GB-hyperhoop} predicts that
the common AH forms even for an arbitrarily long distance between
the two black holes. This obviously contradicts our numerical result
in Fig.~\ref{Z0_AGB_UnitADM}. Therefore, the condition for the AH
formation cannot be obtained at least by a straightforward
extension of the hyperhoop conjecture, and 
a further study is required.

The numerical work in this paper is the first step 
toward simulations of black holes in GB gravity, and 
a lot of extensions can be considered. For example,
it is necessary to extend our method of generating
time-symmetric initial data to 
the method of generating time-asymmetric initial data
like a boosted black hole. For this purpose,
the extension of the Bowen-York method \cite{BY80} should be an
interesting possibility.
Also, the time evolution of black hole initial data
could be done using the approximation analogous
to close-limit or close-slow methods in GR \cite{PP94,YSS05,YSS06}.
The final goal would be to develop numerically stable formulations
of numerical GB gravity by extending $N+1$ formalism and simulate
black hole systems fully numerically to clarify a lot of
interesting phenomena such as time evolution of
an unstable GB black hole, rotating systems,
and high-velocity collision of black holes.

\acknowledgments

The author thanks Hideo Kodama and Hideki Maeda for helpful discussions that
motivated this work. 
This work was supported by the Grant-in-Aid for
Scientific Research (A) (22244030).



\end{document}